# Systematic decay studies of even-even $^{132-138}$Nd, $^{144-158}$Gd, $^{176-196}$Hg and $^{192-198}$Pb isotopes

K. P. Santhosh* and Sabina Sahadevan

*School of Pure and Applied Physics, Kannur University, Payyanur Campus, Payyanur 670 327, INDIA*

**Abstract**

The alpha and cluster decay properties of the $^{132-138}$Nd, $^{144-158}$Gd, $^{176-196}$Hg and $^{192-198}$Pb even–even isotopes in the two mass regions A = 130–158 and A = 180–198 are analysed using the Coulomb and Proximity Potential Model. On examining the clusters at corresponding points in the cold valleys (points with same $A_2$) of the various isotopes of a particular nucleus we find that at certain mass numbers of the parent nuclei, the clusters emitted are getting shifted to the next lower atomic number. It is interesting to see that the change in clusters appears at those isotopes where a change in shape is occurring correspondingly. Such a change of clusters with shape change is studied for the first time in cluster decay. The alpha decay half lives of these nuclei are computed and these are compared with the available experimental alpha decay data. It is seen that the two are in good agreement. On making a comparison of the alpha half lives of the normal deformed and super deformed nuclei, it can be seen that the normal deformed $^{132}$Nd, $^{176-188}$Hg and $^{192}$Pb nuclei are found to be better alpha emitters than the super deformed (in excited state) $^{134,136}$Nd, $^{190-196}$Hg and $^{194}$Pb nuclei. The cluster decay studies reveal that as the atomic number of the parent nuclei increases the N≠Z cluster emissions become equally or more probable than the N=Z emissions. On the whole the alpha and cluster emissions are more probable from the parents in the heavier mass region (A=180-198) than from the parents in the lighter mass region (A= 130-158). The effect of quadrupole ($\beta_2$) and hexadecapole ($\beta_4$) deformations of parent and fragments on half life times are also studied.

*Email address*: drkpsanthosh@gmail.com

## 1 Introduction

The α-decay studies coupled with the cluster decay results have been used to point out the closed shell effects of the nuclei involved in the decay for quite some time now. The Z = 64 sub shell was first predicted by Rasmussen et al. [1] from the study of α - decay energies and Schmidt-Ott et al. [2] noted it from the measurement of the α – decay reduced widths. Gupta and co-workers [3-6] have used the alpha and cluster decay results to predict shell effects in the daughter nuclei. Poenaru et al [7] predicted the spherical $^{100}$Sn daughter radioactivity from their decay studies. Further a deformed radioactivity at Z = 74-76 and N = 98-104 was also noted by Gupta et al. [8] through cluster decay results.

In this paper we have investigated the alpha and cluster decay properties of some even–even isotopes in the two mass regions A = 130–158 and A = 180–198. The nuclei which are studied in the present work are the $^{132-138}$Nd, $^{144-158}$Gd, $^{176-196}$Hg and $^{192-198}$Pb isotopes. All these isotopes are of particular interest for study because of the varied behaviour they present with respect to their shapes. The $150 < A < 190$ mass region to which these isotopes belong is a well known region of deformation having various degrees of deformations and super deformations. There are a number of studies which throw light on the shape variations of all these neutron deficient isotopes. An abrupt change in the mean square charge radius, $\delta\langle r^2\rangle$ was seen in Hg isotopes at mass numbers, $A \leq 186$ from isotope shift measurements [9-11]. The change in the mean square charge radius, $\delta\langle r^2\rangle$ between A = 186 and 185 was interpreted as a shape transition [12-14] while the isomeric shift in $^{185}$Hg was explained in terms of shape coexistence phenomena [15-17], a picture already suggested from nuclear spectroscopic measurements [18-21]. The even-even Hg isotopes with $180 \leq A \leq 188$ exhibit shape coexistence at low spins [19-33]. The coexistence of spherical and deformed degrees of freedom in light lead isotopes was proposed by a number of authors [17, 34-37]. The A = 144-158 mass region of the Gadolinium isotopes is a region in which deformation sets in rapidly and also the effects of the minor closed shell at Z = 64 for $N \leq 88$ have been observed. The even- even Gadolinium isotopes with $86 \leq N \leq 106$ are shown to have positive ground state deformations in the Skyrme HF model [38]. In the case of $^{152}$Gd the admixture of the negative deformation also seems to be important.

The two mass regions under study also have the peculiarity that some of the nuclei among them are super deformed. Note that superdeformation here refers to the observation of super deformed (excited) bands in these nuclei, though their ground-state deformations are not very much different from other neighbouring nuclei. On the other hand, a deformed or normal deformed nucleus is the one where super deformed band(s) are not observed. Super deformed (excited) bands have been experimentally observed in $^{133-137}$Nd, $^{146-150}$Gd, $^{189-194}$Hg and $^{193,194}$Pb isotopes. In the present paper the alpha and cluster decay results have been analysed to get an idea of how the super deformed (in excited state) nuclei behave with respect to alpha and cluster decays as compared to their normal deformed counterparts.

The Coulomb and Proximity Potential Model [39] have been used to study alpha and cluster radioactivity in various mass regions of the nuclear chart. In this model the interacting barrier for the post scission region is taken as the sum of Coulomb and Proximity potential and for the overlap region a simple power law interpolation is used. The model (CPPM) was used to study cluster radioactivity and has predicted half life times for various proton rich parents with Z = 56–64 and N = 56–72, decaying to doubly magic $^{100}$Sn. The model (CPPM) was also used to study the cold valleys in the radioactive decay of $^{248-254}$Cf isotopes [40] and the computed alpha decay half lives are in close agreement with the experimental data. This model is modified [41] by incorporating the ground-state deformations $\beta_2$ and $\beta_4$ of the parent and daughter treating cluster as a sphere and the effect of deformation on the half-lives are studied. In the present paper we have calculated the half lives and other characteristics of the chosen nuclei within the Coulomb and Proximity Potential Model. The effect of quadrupole and hexadecapole deformations of parents and fragments on the half lives is also studied. The details of the model are given in Sec.2. The results obtained and the discussions therein are given in Sec. 3. The study of the effect of deformation on cluster decay half life time is given in Sec. 4. Finally the conclusions from our study are given in Sec. 5.

## 2 The Coulomb and Proximity Potential Model

In Coulomb and Proximity Potential Model (CPPM), potential energy barrier is taken as the sum of Coulomb potential, proximity potential and centrifugal potential for the touching configuration and for the separated fragments. For the pre-scission (overlap) region, simple power law interpolation as done by Shi and Swiatecki [42] is used. The inclusion of proximity potential reduces the height of the potential barrier, which closely agrees with the experimental result. The proximity potential was first used by Shi and Swiatecki [42] in an empirical manner and has been quiet extensively used over a decade by Gupta et al [43] in the preformed cluster model (PCM) which is based on pocket formula of Blocki et al [44] given as:

$$\Phi(\varepsilon) = -\left(\frac{1}{2}\right)(\varepsilon - 2.54)^2 - 0.0852\,(\varepsilon - 2.54)^3 \qquad \text{for } \varepsilon \leq 1.2511 \qquad (1)$$

$$\Phi(\varepsilon) = -3.437 \exp\left(\frac{-\varepsilon}{0.75}\right) \qquad \text{for } \varepsilon \geq 1.2511 \qquad (2)$$

Here $\Phi$ is the universal proximity potential. In the present model, another formulation of proximity potential [45] is used as given by equations (6) and (7). In this model cluster formation probability is taken as unity for all clusters irrespective of their masses, so the present model differs from PCM by a factor $P_0$, the cluster formation probability. But we have included the contribution of both internal (overlap region) and external barrier for penetrability calculations. We would like to mention that within fission model [46-49] penetrability through the internal barrier (overlap region) represent the cluster formation probability. In the present model assault frequency, ν is calculated for each parent-cluster combination which is associated with zero point vibration energy. But Shi and Swiatecki [50] get ν empirically, unrealistic values $10^{22}$ for even A parent and $10^{20}$ for odd A parent.

The interacting potential barrier for a parent nucleus exhibiting exotic decay is given by

$$V = \frac{Z_1 Z_2 e^2}{r} + V_p(z) + \frac{\hbar^2 \ell(\ell+1)}{2\mu r^2} \qquad \text{, for } z > 0 \tag{3}$$

Here $Z_1$ and $Z_2$ are the atomic numbers of the daughter and emitted cluster, 'z' is the distance between the near surfaces of the fragments, 'r' is the distance between fragment centres, $\ell$ the angular momentum, $\mu$ the reduced mass and $V_P$ is the proximity potential given by Blocki et al [44] as

$$V_p(z) = 4\pi\gamma b\left[\frac{C_1 C_2}{(C_1 + C_2)}\right]\Phi\left(\frac{z}{b}\right) \tag{4}$$

With the nuclear surface tension coefficient,

$$\gamma = 0.9517\,[1 - 1.7826\,(N - Z)^2 / A^2] \qquad \text{MeV/fm}^2 \tag{5}$$

Here N, Z and A represent neutron, proton and mass number of parent, $\Phi$ represent the universal proximity potential [45] given as

$$\Phi(\varepsilon) = -4.41 e^{-\varepsilon/0.7176} \quad , \qquad \text{for } \varepsilon \geq 1.9475 \tag{6}$$

$$\Phi(\varepsilon) = -1.7817 + 0.9270\,\varepsilon + 0.0169\,\varepsilon^2 - 0.05148\,\varepsilon^3 \qquad \text{for } 0 \leq \varepsilon \leq 1.9475 \tag{7}$$

With ε = z/b, where the width (diffuseness) of the nuclear surface b ≈1 and Siissmann central radii $C_i$ of fragments related to sharp radii $R_i$ as

$$C_i = R_i - \left(\frac{b^2}{R_i}\right) \tag{8}$$

For $R_i$ we use semi empirical formula in terms of mass number $A_i$ as [44]

$$R_i = 1.28 A_i^{1/3} - 0.76 + 0.8 A_i^{-1/3} \tag{9}$$

Using one dimensional WKB approximation, the barrier penetrability P is given as

$$\mathrm{P} = \exp\left\{ -\frac{2}{\hbar} \int_a^b \sqrt{2\mu (V - Q)}\, dz \right\} \tag{10}$$

Here the mass parameter is replaced by $\mu = mA_1A_2/A$, where m is the nucleon mass and $A_1$, $A_2$ are the mass numbers of daughter and emitted cluster respectively. The turning points "a" and "b" are determined from the equation $V(a) = V(b) = Q$. The above integral can be evaluated numerically or analytically, and the half life time is given by

$$T_{1/2} = \left( \frac{\ln 2}{\lambda} \right) = \left( \frac{\ln 2}{\upsilon P} \right) \tag{11}$$

Where, $\upsilon = \left( \frac{\omega}{2\pi} \right) = \left( \frac{2E_v}{h} \right)$ represent the number of assaults on the barrier per second and λ the decay constant. $E_v$, the empirical zero point vibration energy is given as [51]

$$E_v = Q\left\{ 0.056 + 0.039 \exp\left[ \frac{(4 - A_2)}{2.5} \right] \right\} \quad , \qquad \text{for } A_2 \geq 4 \tag{12}$$

The Coulomb interaction between the two deformed and oriented nuclei taken from [52] with higher multipole deformations included [53, 54] is given as

$$V_C = \frac{Z_1 Z_2 e^2}{r} + 3Z_1 Z_2 e^2 \sum_{\lambda, i=1,2} \frac{1}{2\lambda + 1} \frac{R_{0i}^{\lambda}}{r^{\lambda+1}} Y_{\lambda}^{(0)}(\alpha_i) \times \left[ \beta_{\lambda i} + \frac{4}{7} \beta_{\lambda i}^2 Y_{\lambda}^{(0)}(\alpha_i) \delta_{\lambda,2} \right], \tag{13}$$

with $$R_i(\alpha_i) = R_{0i}\left[ 1 + \sum_{\lambda} \beta_{\lambda i} Y_{\lambda}^{0}(\alpha_i) \right], \tag{14}$$

where $R_{0i} = 1.28 A_i^{1/3} - 0.76 + 0.8 A_i^{-1/3}$. Here $\alpha_i$ is the angle between the radius vector and symmetry axis of the $i^{\text{th}}$ nuclei (see figure 1 of ref. [53]). Note that the quadrupole interaction term proportional to $\beta_{21}\beta_{22}$ is neglected because of its short-range character.

## 3 Results and discussion

We first present the alpha decay results for the $^{132\text{-}138}$Nd, $^{144\text{-}158}$Gd, $^{176\text{-}196}$Hg and $^{192\text{-}198}$Pb isotopes and then analyse the cluster decay calculations. The driving potential (difference between interaction potential and Q value of reaction) of a compound nucleus is calculated for all possible cluster-daughter combinations as a function of mass and charge asymmetries. The Q values are computed using the experimental binding energies of Audi et al [55]. The driving potential versus $A_2$ (mass of one fragment) graphs are plotted for each set of nuclei and are displayed in Figures 1-4. From the plots it can be seen that the potential energy minima occur at $^{4}$He and all the other typical clusters ($^{8,10}$Be, $^{12, 14}$C, $^{16, 18}$O, $^{20, 22}$Ne etc.) for almost all the nuclei. One interesting result that comes up from these plots is the change of clusters at certain mass numbers of the parent nuclei, i.e., when we examine the clusters at corresponding points in the cold valleys (points with same $A_2$) for the various isotopes of a particular parent nucleus we find that at certain mass numbers of the parent nuclei, the clusters emitted are getting shifted to the next lower atomic number. Thus in the cold valley plots (Figure 1) of the Gd isotopes it can be seen that the cluster $^{24}$Mg in the cold valley plot of $^{144}$Gd changes to $^{24}$Ne in the cold valley of $^{146}$Gd. Similarly $^{38}$Ar in the cold valley of $^{144}$Gd changes to $^{38}$S in the valley of $^{146}$Gd, again $^{44}$Ca becomes $^{44}$Ar and this goes on up to $^{70}$Ge at $^{144}$Gd which is changed to $^{70}$Zn in the cold valley of $^{146}$Gd. The cold valley plots for Nd isotopes (Figure 2) show a similar trend with changes in clusters emerging at $^{134}$Nd and $^{138}$Nd. The cluster $^{24}$Mg in the plot of $^{132}$Nd changes to $^{24}$Ne at $^{134}$Nd, then $^{28}$Si goes on to become $^{28}$Mg and this continues till $^{64}$Zn at $^{132}$Nd which is getting changed to $^{64}$Ni in the cold valley of $^{134}$Nd. Similarly the clusters $^{20}$Ne, $^{34}$S, $^{48}$Ti, $^{56}$Fe and $^{66}$Zn in the cold valley of $^{136}$Nd are changed to $^{20}$O, $^{34}$Si, $^{48}$Ca, $^{56}$Cr and $^{66}$Ni in the valley of $^{138}$Nd. Moving on to the cold valley plots (Figure 3) of the Hg isotopes changes in clusters emerges at $^{186}$Hg and $^{196}$Hg. The minima at $^{20}$Ne, $^{34}$S, $^{48}$Ti, $^{56}$Fe, $^{66}$Zn and $^{72}$Ge in the valley of $^{184}$Hg are getting shifted down to $^{20}$O, $^{34}$Si, $^{48}$Ca, $^{56}$Cr, $^{66}$Ni and $^{72}$Zn in the plot of $^{186}$Hg. In the cold valley for $^{196}$Hg it is found that cluster changes are emerging at $^{22}$Ne, $^{26}$Mg, $^{60}$Fe and $^{70}$Zn i.e, $^{22}$Ne in the valley of $^{194}$Hg changes to $^{22}$O in the plot of $^{196}$Hg. Similarly $^{26}$Mg, $^{60}$Fe and $^{70}$Zn in the valley of $^{194}$Hg changes to $^{26}$Ne, $^{60}$Cr and $^{70}$Ni in the valley of $^{196}$Hg. In the case of the Pb parents there are changes at $^{198}$Pb but it is not a continuous one as can be seen at the other cases discussed above.

The change in clusters appears at those nuclei where a change in shape is occurring correspondingly. As mentioned in the earlier paragraph the first change in clusters is observed at $^{146}$Gd. According to the tables of Möller et al. [56] the quadrapole deformation parameters $\beta_2$ of $^{145}$Gd and $^{146}$Gd are -0.053 and 0 respectively, thereby indicating a change in shape from oblate to spherical at $^{146}$Gd. For the low spin states in the neutron deficient Gd isotopes the deformation decreases with increasing neutron number and a spherical shell structure develops when the shell closure at N = 82 is approached. More precisely in $^{142}$Gd a ground band is observed which has a structure resembling that of a vibrator or a rotor with a modest deformation. In contrast, $^{144}$Gd [57] show a structure which requires an explanation in terms of shell model excitations. This clearly indicates a shape change at the $^{146}$Gd isotope from deformed to spherical. Thus cluster changes are taking place at $^{146}$Gd where correspondingly a shape change is observed. Our studies show cluster changes at $^{134}$Nd where there is a shape variation from normal deformed to superdeformed (in excited state). The $^{132}$Nd nucleus with experimental quadrupole deformation, $\beta_2 = 0.349$ [58] is normal deformed prolate while the $^{134}$Nd nucleus is superdeformed (in excited state). The next change in clusters is found at $^{138}$Nd. A shape change from prolate ($\beta_2 = 0.154$ [56]) to oblate ($\beta_2 = -0.138$ [56]) is taking place when we go from $^{137}$Nd to $^{138}$Nd. Experimentally it is also found that $^{137}$Nd is superdeformed (in excited state) therefore there is also a change from superdeformed to normal deformed when going from$^{137}$Nd to $^{138}$Nd. Moving on to the Hg isotopes changes in clusters are observed at $^{186}$Hg and $^{196}$Hg. In the case of $^{186}$Hg we find shape variations from oblate to prolate i.e, $^{185}$Hg has $\beta_2 = -0.139$ [56] while $^{186}$Hg has $\beta_{2expt.} = 0.132$ [58]. Similarly the oblate shape at $^{195}$Hg ($\beta_2 = -0.130$ [56]) changes to prolate at $^{196}$Hg ($\beta_{2expt.} = 0.1155$ [58]). Also at $^{196}$Hg there is a transition from superdeformed (in excited state) to normal deformed because $^{194}$Hg has superdeformed excited bands while $^{195,196}$Hg nuclei are normal deformed. In the case of Pb isotopes no appreciable change in clusters is observed in our studies which mean there is no shape variation when going from $^{192}$Pb to $^{198}$Pb. Recent experiments on the ground-state charge mean-square radii also show that the ground-state in all even-even Pb nuclei remains of spherical shape. Thus according to these observations, whenever a systematic change in clusters occurs at a particular isotope, a corresponding shape transition is occurring there. It is evident that our ground state cluster decay model is able to predict the shape change of nuclei in the ground state as well as in the excited state.

### 3.1 Alpha decay results

The $\log_{10}(T_{1/2})$ values corresponding to alpha decay of all the chosen Nd-Pb parent nuclei are plotted against the mass number of the parent and is presented in Figure 5. All the known experimental alpha decay half lives are included in the figure and it can be seen that they match well with our predictions. The experimental alpha decay data are from the refs. [59-68]. In the case of the Nd isotopes $^{144}$Nd is the only nucleus around the chosen range which has experimental alpha decay data therefore its half life has been included in the graph. Along the same line the alpha decay calculations of $^{149}$Gd, $^{151}$Gd and $^{177,179,181,183,185}$Hg have been performed and included in the $\log_{10}(T_{1/2})$ graph for alpha decay. As far as the Pb isotopes are concerned the alpha half lives of all the nuclei (both even-even and even-odd) in the range A = 182-191 and again the alpha half lives of the even-even isotopes from A = 200 to A = 210 have been estimated and presented in the alpha decay half lives graph. This has been done so because all the nuclei in the range A = 182-191 have experimental data while calculations for the isotopes from A = 200 to A = 210 have been conducted to include the lone experimental data of $^{210}$Pb.

From the alpha decay half lives graph it is seen that $^{132\ \&144}$Nd and $^{148,149,150,151,152}$Gd have their half lives within the present experimental limit of $10^{30}$s. Again with the exception of $^{196}$Hg all the other Hg nuclei considered (both even-even and even-odd) have their alpha decay half lives within the present experimental limit of $10^{30}$s. Similarly all the lead isotopes in the range 182 < Z < 200 (both even-even and even-odd) and the $^{210}$Pb nucleus have alpha decay half lives within the order of $10^{30}$s. This means that all these nuclei are alpha instable and most of them have been experimentally detected as can be seen from the figure. Again from a first glance at figure 5 it is obvious that the elements in the heavier mass region (A = 180-198) are more prone to alpha decay than the ones in the lighter mass region (A = 130-158).

Figure 5 for the alpha decay results also presents a number of shell structure effects of both the parent and the daughter nuclei. The low $T_{1/2}$ value for the $^{144}$Nd nucleus is due to the N = 82 closure in the $^{140}$Ce daughter. The extremely large half life of the $^{146}$Gd parent is attributed to the N = 82 spherical shell closure while the instability of $^{148,149,150,151,152}$Gd parents are due to N = 82 and N ≈ 82 shell closure in the corresponding Sm daughter nuclei. The lowest alpha decay half lives amongst all the nuclei are got for the $^{176\text{-}188}$Hg ($T_{1/2}$ = 0.059s for $^{176}$Hg and $T_{1/2}$ = $2.44\times10^{9}$s for $^{188}$Hg) and the $^{182\text{-}190}$Pb parent nuclei ($T_{1/2}$ = 0.119s for $^{182}$Pb and $T_{1/2}$ = $6.8\times10^{4}$s for $^{190}$Pb). The daughter nuclei for the alpha decay of the $^{176\text{-}184}$Hg parents are the Pt isotopes with Z ≈ 76 and N = 94, 96, 98, 100, 102, 104 and 106 which implies the magic or semi-magic nature of these neutron shells. The N = 98, 100, 102, 104 and 106 shell closures are again reflected in the Z ≈ 82 Hg daughter nuclei for the alpha decay of the $^{182\text{-}190}$Pb parents. Several authors [3, 8, 69], have predicted Z = 76, 74 and N = 96, 98, 100, 102 and 104 as major (deformed) closed shells. Also the Z = 76 shell gap is got as a clear dip in the experimental shell energy curves [70] for Sr and Zr isotopes. Then there is the very obvious extremely large rise in half life at the doubly magic $^{208}$Pb parent nucleus (Z = 82 and N = 126) and again a sudden dip at the $^{210}$Pb parent which is of course due to the Z ≈ 82 and N = 126 shell closures in the $^{206}$Hg daughter.

On taking a comparison of the normal deformed and super deformed (in excited state) nuclei, it can be seen that except for the case of the Gd isotopes for all the other three set of nuclei, the normal deformed nuclei ( $^{132}$Nd, $^{176-188}$Hg and $^{192}$Pb nuclei ) are better alpha emitters than the super deformed ones ($^{134,136}$Nd, $^{190-194}$Hg and $^{194}$Pb nuclei ). In the case of the Gd isotopes the alpha decay half lives get influenced by the shell closure effects of the nuclei involved in the decay process. The super deformed $^{146}$Gd nucleus is stable for alpha emission with a half life of $4.69\times10^{105}$s while the super deformed $^{148,149,150}$Gd nuclei are alpha instable with half lives $2.05\times10^{10}$s, $8.02\times10^{11}$s and $9.09\times10^{14}$s respectively.

**3.2 Cluster decay results**

The N=Z cluster and N≠Z cluster results are studied separately for each set of parent nuclei. The $\log_{10}(T_{1/2})$ values for the N=Z clusters and N≠Z clusters are plotted against the mass number of the parents and are displayed in Figures 6-13; the corresponding observations for each parent are presented below. The graphs for Nd parents show that there are almost equal number of N=Z and N≠Z cluster emissions (six N=Z clusters and eight N≠Z clusters) from them while in the case of the Gd parents there are more N≠Z clusters being emitted than the N=Z clusters (five N=Z clusters and nine N≠Z clusters). When we analyze the corresponding half life values we can see that for the Nd parents except for few emissions ($^{12}$C, $^{16}$O, $^{28}$Si, $^{30}$Si, $^{34}$S from $^{132}$Nd) almost all the cluster decays (both N=Z and N≠Z) have their $T_{1/2} > 10^{50}$s and for the Gd parents except for two or three N=Z clusters ($^{12}$C from $^{150,152}$Gd and $^{16}$O from $^{154}$Gd) all the other cluster decays have their $T_{1/2} > 10^{50}$s. On comparing the cluster emissions from the Hg and Pb parents we find that there are equal numbers of N=Z and N≠Z clusters (seven N=Z and seven N≠Z clusters) getting emitted from Hg parents while there are more number of N≠Z cluster emissions than N=Z emissions (three N=Z clusters and eleven N≠Z clusters) from Pb parents. On making a study of the half lives of the cluster decays from Hg and Pb parents we can see that more than half the decays (both N=Z and N≠Z) from the Hg and Pb parents have their $T_{1/2} > 10^{50}$s. Thus we may infer that as the N:Z ratio of the parent nuclei increases the number of N≠Z clusters emitted are becoming equal to or greater than the number of N=Z clusters. Again from the analysis of the cluster decay half lives it is clear that the cluster decay rates for the parents in the heavier mass region are larger (smaller cluster decay half lives) than those for the parents in the lighter mass region. Thus the parents in the lighter mass region (A= 130-158) are found to be more stable against cluster decays than the parents in the heavier mass region (A=180-198). In the following sections we will discuss these results in detail for each set of parent nuclei separately.

**3.2.1 Nd parents**

Figures 6 and 7 give the $\log_{10}(T_{1/2})$ values plotted as a function of the parent mass number for the N = Z clusters and N ≠ Z clusters respectively. The $^{132}$Nd parent is weakly stable against $^{12}$C, $^{16}$O (N = Z clusters) and $^{30}$Si, $^{34}$S (N ≠ Z clusters) decays. Amongst these $^{16}$O cluster has the lowest half life of 1.68 x $10^{46}$s. The reason for the $^{16}$O decay from $^{132}$Nd being the most probable among the Nd parents is due to the proximity to the Z = 50 magic shell coupled with the midshell effect of the N = 64 neutron shell in the $^{116}$Te daughter. Interestingly the $^{30}$Si and $^{34}$S heavy cluster decays from $^{132}$Nd have their half lives (1.84 x $10^{47}$s and 2.96 x $10^{49}$s respectively) lower than the $^{12}$C light cluster decay (1.15 x $10^{50}$ s). It must be the stabilising effects of the N = 56 semi magic neutron number [70] which is the reason for the low half lives of the $^{30}$Si and $^{34}$S decays. The daughter nuclei for the $^{30}$Si and $^{34}$S decays are the $^{102}$Pd and $^{98}$Ru nuclei with neutron numbers N = 56 and N = 54.

The $^{8}$Be decay from $^{134-138}$Nd as well as the $^{14}$C decay from $^{132-138}$Nd has their half lives, $T_{1/2} > 10^{100}$s and therefore these cluster emissions must be considered as being hindered for the chosen Nd parents. Also from Figures 6 and 7 it can be seen that for the $^{136}$Nd parent the $^{16}$O, $^{30}$Si decay half lives are comparable with the alpha decay half life while for the $^{138}$Nd parent the $^{12}$C, $^{16}$O, $^{22}$Ne, $^{26}$Mg, $^{28}$Mg, $^{30}$Si, $^{32}$Si,$^{34}$Si and $^{36}$S decays are more probable than the alpha decay. However the decay half lives in all these cases is very large, $T_{1/2} > 10^{50}$s. Thus, although a number of shell effects are evident from the cluster decay results, the Nd parents are, on the whole, very stable against cluster decays.

**3.2.2 Gd parents**

The $\log_{10}(T_{1/2})$ results of Gd parents for N = Z clusters are displayed in Figure 8 and those for N ≠ Z clusters in Figure 9. The $^{12}$C cluster from $^{152}$Gd parent has the lowest half life ($T_{1/2}$=2.44 x $10^{41}$ s) among all the Gd parents. This low half life value of the $^{12}$C decay from $^{152}$Gd parent is due to the N = 82 spherical shell closure in the $^{140}$Ce daughter. Then there are dips for the $^{8}$Be and $^{16}$O emission from the $^{150}$Gd and $^{154}$Gd parents respectively. $^{8}$Be emission from $^{150}$Gd isotope is due to the spherical N = 82 shell closure of the daughter $^{142}$Nd while the

$^{16}$O decay from $^{154}$Gd is due to N = 82 spherical shell closure in the $^{140}$Ce daughter. The N = 82 spherical shell closure of the $^{146}$Gd parent is obtained as a rise in $T_{1/2}$ values for all clusters (except $^{20}$Ne) at the $^{146}$Gd isotope. Of all these clusters only $^{8}$Be from $^{150}$Gd and $^{12}$C cluster from $^{150}$Gd and $^{152}$Gd can be classified as weakly stable as all the other decays have relatively large half lives.

Coming to the N ≠ Z clusters, all the decays are either stable or very stable with the lowest half lives for the $^{30}$Si decay from $^{144}$Gd and the $^{14}$C decay from $^{154}$Gd. The low value for the $^{30}$Si decay is due to the Z = 50 spherical shell closure in the $^{114}$Sn daughter while the $^{14}$C decay is bringing forth the N = 82 shell closure in the $^{140}$Ce daughter which is also evident as a dip in the $\log_{10}(T_{1/2})$ versus mass number of parent graph. Another dip is got at the $^{156}$Gd isotope for the $^{18}$O decay. This is again due to the N = 82 shell closure in the $^{138}$Ba daughter. The Sn daughter radioactivity is again coming up as a small dip at the $^{154}$Gd parent for the $^{32}$Si decay. It has also one of the lowest half lives ($T_{1/2}$=3.33x10$^{61}$s) among the N ≠ Z cluster decays. Although these decays have half lives much larger than the present experimental limit they are many orders smaller than the before mentioned N = Z cluster decays except for two or three cases.

As can be seen from the $\log_{10}(T_{1/2})$ graphs there is a dip at the $^{154}$Gd nucleus for a large number of clusters with the most probable one being the $^{16}$O decay with a half life of 8.04 x 10$^{50}$s. This decrease in half life values at the $^{154}$Gd nucleus is particularly visible in the case of the N ≠ Z clusters wherein it can be seen that the isotopes on the left and right side of the $^{154}$Gd nucleus are having larger $T_{1/2}$ values than $^{154}$Gd for almost all the cluster emissions. Thus the $^{154}$Gd nucleus is found to be less stable against cluster decays as compared to its neighbouring isotopes. This decrease in half lives at $^{154}$Gd is a classic case of daughter shell effects coming into play – almost all the dips in half lives at $^{154}$Gd is due to the shell closures in the corresponding daughter nuclei and they have already been discussed in the earlier paragraphs.

### 3.2.3 Hg parents

Figure 10 gives the $\log_{10}(T_{1/2})$ results for the N = Z clusters while Figure 11 gives the same for the N ≠ Z clusters. Of the probable N = Z clusters $^{12}$C from $^{176\text{-}184}$Hg, $^{8}$Be from $^{176\text{-}182}$Hg, $^{16}$O from $^{176\text{-}180}$Hg, $^{28}$Si from $^{176}$Hg and $^{24}$Mg from $^{176}$Hg parents have half lives within the present measurable limit of 10$^{30}$s. Also the decay of $^{12}$C from $^{186}$Hg, $^{8}$Be from $^{184}$Hg, $^{16}$O from $^{182}$Hg and $^{20}$Ne from $^{176}$Hg have half lives very close to 10$^{30}$s. Since all the so far observed daughters in decays of radioactive nuclei were magic or nearly magic nuclei the $^{12}$C emission from $^{176\text{-}184}$Hg, $^{8}$Be from $^{176\text{-}182}$Hg and $^{16}$O from $^{176\text{-}180}$Hg indicate shell closures at $^{164\text{-}172}$W, $^{168\text{-}174}$Os and $^{160\text{-}164}$Hf daughter nuclei. That is the Z = 76, 74, 72 proton shell closures and the N = 88, 90, 92, 94, 96, 98 neutron shell closures are coming up at these nuclei. This means predicting major closed shells at Z = 72 and N = 88, 90, 92 and 94 in addition to the already predicted Z = 76, 74 and N = 96-104 shell closures. Coming to the N ≠ Z clusters, from figure 11 it can be seen that $^{30}$Si emission from $^{176}$Hg and $^{178}$Hg and $^{26}$Mg from $^{176}$Hg have half live values within the present experimental limit. The clusters $^{30}$Si and $^{34}$S from $^{180}$Hg also decay with half lives very close to the present experimental limit.

As expected, the $\log_{10}(T_{1/2})$ versus mass number of parent graphs is showing dips at those masses whose decay involves daughters with known spherical magicities. Thus in the case of Hg parents, the minima occur at $^{180}$Hg parent for the decay of $^{32}$Si cluster with daughter $^{148}$Dy having N=82 shell closure; then there are dips at $^{178}$Hg parent for $^{30}$Si cluster decay, $^{180}$Hg parent for $^{34}$S cluster decay and $^{182}$Hg parent for $^{36}$S cluster decay, all of which have daughter nuclei ($^{148}$Dy, $^{146}$Gd and $^{146}$Gd respectively) with N=82 spherical shell closure. This is once again proof for the role of shell closure in cluster decay.

The $^{180}$Hg and $^{186}$Hg isotopes with just two protons less than the Z = 82 shell closure together with the proximity to the N = 102 and N = 108 neutron shell closures (both nuclei just two neutrons less than the N = 102 and N = 108 neutron shell closures respectively) are to be considered as stable nuclei. A strongly prolate neutron stable shell has been predicted at N = 102 by Hannachi et al., [33] while the shell closure at N = 108 has been emphasised by Bengtsson et al [70]. The analysis of the data on γ-ray spectra by Carpenter et al., [71] has also revealed a prolate minimum for the ground state at N = 102. The decay results for the Hg isotopes present some interesting cases of instabilities at these stable nuclei – these are the $^{30}$Si and $^{34}$S decays from $^{180}$Hg and the $^{8}$Be and $^{12}$C decays from $^{186}$Hg. All these decays have half lives within or very close to the present experimental limit. The stability of the $^{150}$Dy daughter nucleus (with N ≈ 82 and Z value exactly in the middle of two spherical magic numbers 50 and 82) is brought forth in the $^{30}$Si decay from $^{180}$Hg while the $^{34}$S decay from $^{180}$Hg is due to the stability of the $^{146}$Gd daughter nucleus. On a similar note the Z = 74, 76 and N = 100, 102 shell closures are again getting established from the $^{174}$W and $^{178}$Os deformed daughter radio activities in the $^{8}$Be and $^{12}$C decay from $^{186}$Hg. Gupta et al.[3] have observed similar cases of instabilities from $^{120}$Ba and $^{186}$Hg nuclei.

### 3.2.4 Pb parents

Figure 12 which is the plot of $\log_{10}(T_{1/2})$ versus A, mass number of parent for N = Z clusters shows that $^{12}$C cluster decay has the lowest half life among all cluster emissions from Pb parents with $^{12}$C from $^{192}$Pb having a half life of $5\times10^{33}$s which is close to the present experimental limit of $10^{30}$s. The $^{8}$Be emission from $^{192,194}$Pb and $^{16}$O emission from $^{192\text{-}196}$Pb have their half lives in the range, $10^{37}\text{s} < T_{1/2} < 10^{50}\text{s}$ and therefore they can be classified as being weakly stable. Coming to the N ≠ Z clusters (Figure 13) $^{30}$Si cluster from $^{192}$Pb is weakly stable with the lowest half life ($T_{1/2} = 10^{45}$s). Also $^{32}$Si from $^{192,194}$Pb, $^{22}$Ne and $^{36}$S from $^{192}$Pb and $^{26}$Mg from $^{192,194}$Pb with half lives in the range, $10^{47}\text{s} \leq T_{1/2} \leq 10^{50}\text{s}$ are all weakly stable.

The lowest $T_{1/2}$ values of $^{12}$C emissions from $^{192,194}$Pb is attributed to the Z = 76 proton shell closure coupled with the N = 104, 106 neutron closures in the $^{180,\,182}$Os daughter nuclei. Similarly the low half life value for the $^{8}$Be emission from $^{192}$Pb is due to the N = 106 neutron shell closure in the $^{184}$Pt daughter nucleus while the $^{16}$O emissions from the $^{192,194,196}$Pb are due to the Z = 74 proton shell closure together with the N = 102, 104, 106 neutron shell closures in the $^{176,178,180}$W daughter nuclei.

Amongst the N ≠ Z clusters the $^{30}$Si and the $^{32}$Si decays from $^{192,194}$Pb parents have the lowest half lives. The daughter nuclei for the $^{30}$Si decays are the $^{162,164}$Er nuclei and the daughters for the $^{32}$Si decays are the $^{160,162}$Er nuclei respectively. Thus it can be seen that these emissions are occurring due to the N = 92, 94, 96 neutron shell closures coupled with the mid-shell effect of the Z = 68 proton shell in the Er isotopes. The $^{22}$Ne emission from $^{192}$Pb has $^{170}$Hf as daughter and $^{36}$S emission from $^{192}$Pb has $^{156}$Dy as the daughter. $^{170}$Hf has Z = 72 and N = 98 while $^{156}$Dy has Z = 66 and N = 90 shell closures in them. Finally the $^{26}$Mg decay from $^{192,194}$Pb has the $^{166,168}$Yb nuclei as the daughters. The N = 96, 98 neutron shell closures are being manifested in the $^{166,168}$Yb daughter nuclei. Thus as in the case of the Hg parents the already predicted Z = 76, 74 and N = 96 -104 shell closures are again coming up at the Pb parents and in addition new shell closures at Z = 72 and N = 90, 92, 94, 106 are also coming up in the daughter nuclei.

## 4 Effect of deformation

In the present work the nuclear proximity potential for oriented and deformed (with higher multipole deformation) nuclei are done following the prescription of Gupta and co-workers [53] with universal proximity potential given in eqns. 6 and 7. The coulomb potential for the two deformed and oriented nuclei is computed using eqn. 13. In fission and cluster decay the fragments are strongly polarized due to nuclear force and accordingly their symmetry axes are aligned. In the present calculations we consider only pole to pole configuration. The proper inclusion of higher multipole deformations along with generalized orientation contributions may prove important in deciding the cluster decay paths of various clusters.

Figures 14-18 and Table 1 represent the comparison of computed logarithm of half life times for the cluster emissions from $^{132-138}$Nd, $^{144-158}$Gd, $^{176-196}$Hg and $^{192-198}$Pb parents for the cases of without deformation (spherical), with quadrupole deformation ($\beta_2$) and with quadrupole and hexadecapole deformations ($\beta_2$ & $\beta_4$). The experimental quadrupole deformations are taken from [58] and the rest from the tables of Moller et al [56]. It is obvious from the table and figures that the half lives decrease with the inclusion of quadrupole deformation $\beta_2$ due to the fact is that it reduces the height and width of the barrier (increases the barrier penetrability). We would like to mention that the sign of hexadecapole deformation have no influence on half life time.

## 5 Conclusions

An analysis of the alpha and cluster decay of the $^{144-158}$Gd, $^{132-138}$Nd, $^{176-196}$Hg and $^{192-198}$Pb even – even isotopes are carried out using the Coulomb and Proximity potential model. This set of parents comprise of both normal deformed and superdeformed (in excited state) nuclei. Therefore one aim of our study was to see how the superdeformed (in excited state) nuclei behave with respect to alpha and cluster decays as compared to their normal deformed counterparts.

One important aspect of these nuclei is the shape variations they present as we move from one isotope to another and therefore their decay results were analyzed to see whether the shape changes influence their decay pattern in any manner. It was found that at certain mass numbers of the parent nuclei, the clusters emitted are getting changed from the ones that are emitted from the isotope just before it. More precisely, when we examine the clusters at corresponding points in the cold valleys (points with same $A_2$) of the various isotopes of a particular parent nucleus we find that at certain mass numbers of the parent nuclei, the clusters emitted are getting shifted to the next lower atomic number and interestingly the change in clusters appears at some of those nuclei where a change in shape is occurring correspondingly. As already mentioned the first change in clusters is observed at $^{146}$Gd and correspondingly there is a change in shape from oblate to spherical at $^{146}$Gd. In the case of Nd isotopes the cluster changes are at $^{134}$Nd and $^{138}$Nd. At $^{134}$Nd there is a shape variation from normal deformed to superdeformed (in excited state) and a shape change from prolate to oblate is taking place at $^{138}$Nd. Here also we would like to mention a change from superdeformed (in excited state) to normal deformed at $^{138}$Nd. Finally in the case of Hg parents there are changes in clusters at $^{186}$Hg and $^{196}$Hg. At $^{186}$Hg we find shape variations from oblate to prolate and similarly at $^{196}$Hg the shape changes from oblate to prolate. Again we would like to mention that at $^{196}$Hg there is a transition from superdeformed (in excited state) to normal deformed. In the case of Pb isotopes no appreciable change in clusters is observed in our studies which mean there is no shape variation when going from $^{192}$Pb to $^{198}$Pb.

The effects of quadrupole and hexadecapole deformations of both parent and fragments on half life times are studied using Coulomb and proximity potential for oriented and deformed nuclei as the interacting barrier. It is found that inclusion of quadrupole deformation $\beta_2$ reduces the height and shape of the barrier (increases barrier penetrability) and hence the half life time decreases.

The most important part of our analysis was regarding the coming up of new shell closures (spherical or deformed) in either of the two mass regions (A = 130 – 158 and A = 180 – 198) under study. It is established that a large decay half life indicates that a parent nucleus is stable as far as alpha and cluster decay is concerned while a small half life is the indication for the daughter to be stable. Based on this idea we have come across a number of shell closures (some known and some new) in the different nuclei involved in the study.

The elements in the heavier mass region (A = 180 - 198) are more prone to alpha decay than the ones in the lighter mass region (A = 130 - 158). The rise in the alpha decay half lives at the $^{146}$Gd and $^{208}$Pb parents is attributed to the spherical sub-magic Z = 64, magic N = 82 and doubly magic Z = 82, N = 126 shell closures respectively. The N=82 spherical shell closure in $^{140}$Ce and $^{144}$Sm daughter nuclei are got as dips in the alpha half lives of the $^{144}$Nd and $^{148}$Gd parent nuclei respectively. The normal deformed $^{132}$Nd, $^{176-188}$Hg and $^{192}$Pb nuclei are found to be better alpha emitters than the super deformed (in excited state) $^{134, 136}$Nd, $^{190-194}$Hg and $^{194}$Pb nuclei.

The cluster decay results show that on the whole cluster emissions are more probable from the parents in the heavier mass region (A=180-198) than from the parents in the lighter mass region (A= 130-158). The N = 82 spherical shell closure of the $^{146}$Gd parent is obtained as a rise in $T_{1/2}$ values for all cluster emissions. The N = 82 spherical shell closure in the daughter nuclei is again getting manifested as dips in the cluster half lives for the cluster emissions from the Gd and Hg parents. The $^{8}$Be emission from $^{176-182}$Hg, $^{12}$C emission from $^{176-184}$Hg, and $^{16}$O emission from $^{176-180}$Hg have their half lives within the present measurable limit of $10^{30}$s thereby indicating the possibility of major (deformed) closed shells at Z = 76, 74, 72 and N = 88, 90, 92, 94, 96 and 98 in their $^{164-172}$W, $^{168-174}$Os and $^{160-164}$Hf daughter nuclei. Similarly the $^{8}$Be emission from $^{192,194}$Pb, the $^{12}$C emission from $^{192,194}$Pb and $^{16}$O emission from $^{192-196}$Pb have the lowest half lives among all the cluster emissions from the Pb parents thereby bringing forth the Z = 76, 74 and N = 102, 104, 106 shell closures again. The $^{30}$Si and $^{34}$S decays from $^{180,186}$Hg isotopes present some interesting cases of instabilities in stable nuclei. The $^{180}$Hg and $^{186}$Hg isotopes with just two protons less than the Z = 82 shell closure together with the proximity to the N = 102 and N = 108 neutron shell closures are stable nuclei but the above mentioned decays from these nuclei have half lives within or very close to the present experimental limit.

**Table 1.** The comparison of computed values of logarithm of half life times for various clusters from $^{176-180}Hg$ parents with and without including deformations.

| Parent nuclei | Emitted cluster | Daughter nuclei | Q value (MeV) | $log_{10}(T_{1/2})$ | | |
|---|---|---|---|---|---|---|
| | | | | Without deformation | With $\beta_2$ | With $\beta_2$ & $\beta_4$ |
| $^{176}Hg$ | $^{4}He$ | $^{172}Pt$ | 6.925 | -1.23 | -2.30 | -2.16 |
| | $^{8}Be$ | $^{168}Os$ | 13.298 | 19.78 | 17.29 | 17.52 |
| | $^{16}O$ | $^{160}Hf$ | 38.927 | 21.99 | 6.90 | 9.45 |
| | $^{18}O$ | $^{158}Hf$ | 31.312 | 41.98 | 27.37 | 29.64 |
| | $^{24}Mg$ | $^{152}Er$ | 62.683 | 29.12 | 6.31 | 8.73 |
| | $^{30}Si$ | $^{146}Dy$ | 75.383 | 29.23 | 12.70 | 12.74 |
| | $^{32}S$ | $^{144}Gd$ | 69.161 | 64.88 | 68.12 | 69.15 |
| | $^{34}S$ | $^{142}Gd$ | 85.112 | 33.70 | 30.17 | 34.34 |
| | $^{36}S$ | $^{140}Gd$ | 80.444 | 41.98 | 15.32 | 20.24 |
| $^{178}Hg$ | $^{4}He$ | $^{174}Pt$ | 6.578 | 0.06 | -1.25 | -1.15 |
| | $^{8}Be$ | $^{170}Os$ | 12.670 | 22.16 | 19.53 | 20.02 |
| | $^{14}C$ | $^{164}W$ | 18.867 | 45.15 | 32.25 | 31.83 |
| | $^{16}O$ | $^{162}Hf$ | 37.594 | 24.63 | 9.24 | 12.55 |
| | $^{18}O$ | $^{160}Hf$ | 30.369 | 44.63 | 28.41 | 30.39 |
| | $^{28}Mg$ | $^{150}Er$ | 56.666 | 40.95 | 19.76 | 28.48 |
| | $^{30}Si$ | $^{148}Dy$ | 75.940 | 27.75 | 11.27 | 11.30 |
| | $^{32}Si$ | $^{146}Dy$ | 70.428 | 37.85 | 26.29 | 26.42 |
| | $^{34}S$ | $^{144}Gd$ | 85.529 | 32.40 | 17.17 | 17.26 |
| | $^{36}S$ | $^{142}Gd$ | 81.241 | 39.90 | 41.63 | 45.82 |
| $^{180}Hg$ | $^{4}He$ | $^{176}Pt$ | 6.266 | 1.31 | -0.33 | -0.26 |
| | $^{8}Be$ | $^{172}Os$ | 12.059 | 24.65 | 21.06 | 21.44 |
| | $^{14}C$ | $^{166}W$ | 18.690 | 45.76 | 32.32 | 32.49 |
| | $^{16}O$ | $^{164}Hf$ | 36.310 | 27.31 | 10.66 | 14.15 |
| | $^{18}O$ | $^{162}Hf$ | 29.770 | 46.31 | 29.87 | 32.60 |
| | $^{28}Mg$ | $^{152}Er$ | 55.290 | 43.71 | 23.04 | 31.96 |
| | $^{30}Si$ | $^{150}Dy$ | 73.560 | 31.67 | 15.26 | 15.30 |
| | $^{32}Si$ | $^{148}Dy$ | 71.720 | 34.84 | 23.21 | 23.31 |
| | $^{34}S$ | $^{146}Gd$ | 85.830 | 31.34 | 16.12 | 16.20 |
| | $^{36}S$ | $^{144}Gd$ | 82.390 | 37.17 | 27.19 | 27.37 |

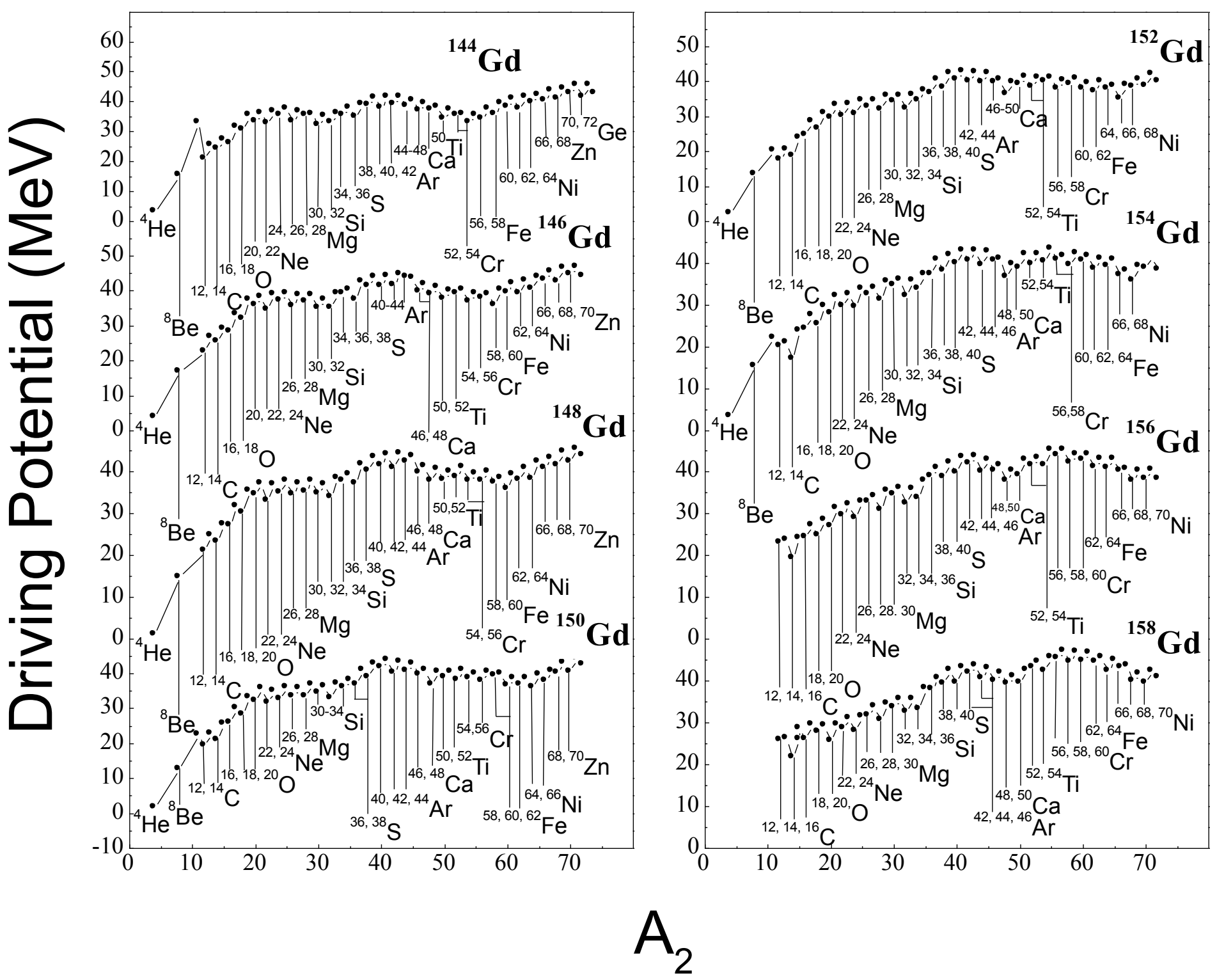

**Fig. 1.** The driving potentials as a function of the mass number of light fragments ($A_2$) for Gd isotopes.

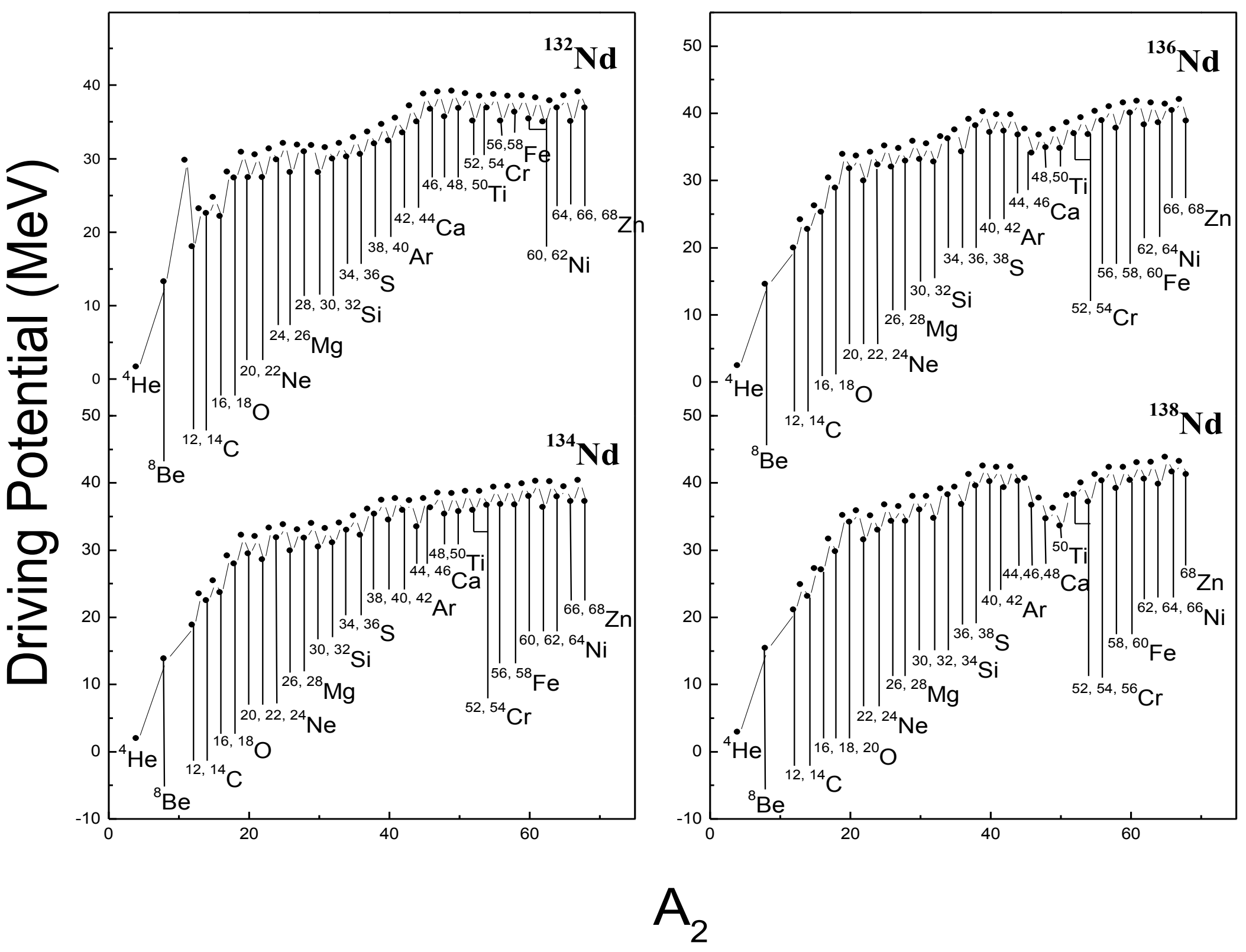

**Fig. 2.** The same as for Fig.1 but for Nd isotopes

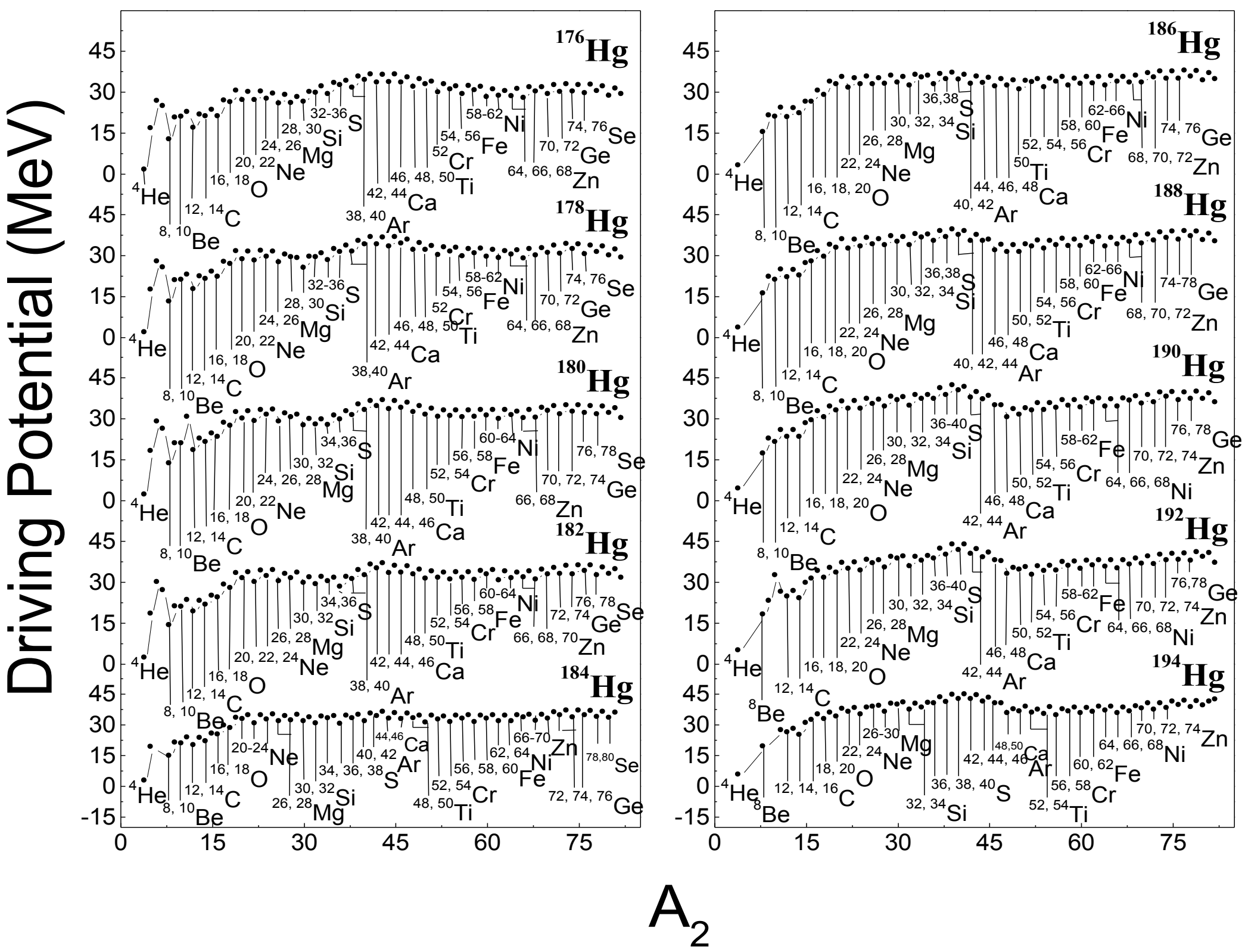

**Fig. 3.** The same as for Fig.1 but for Hg isotopes

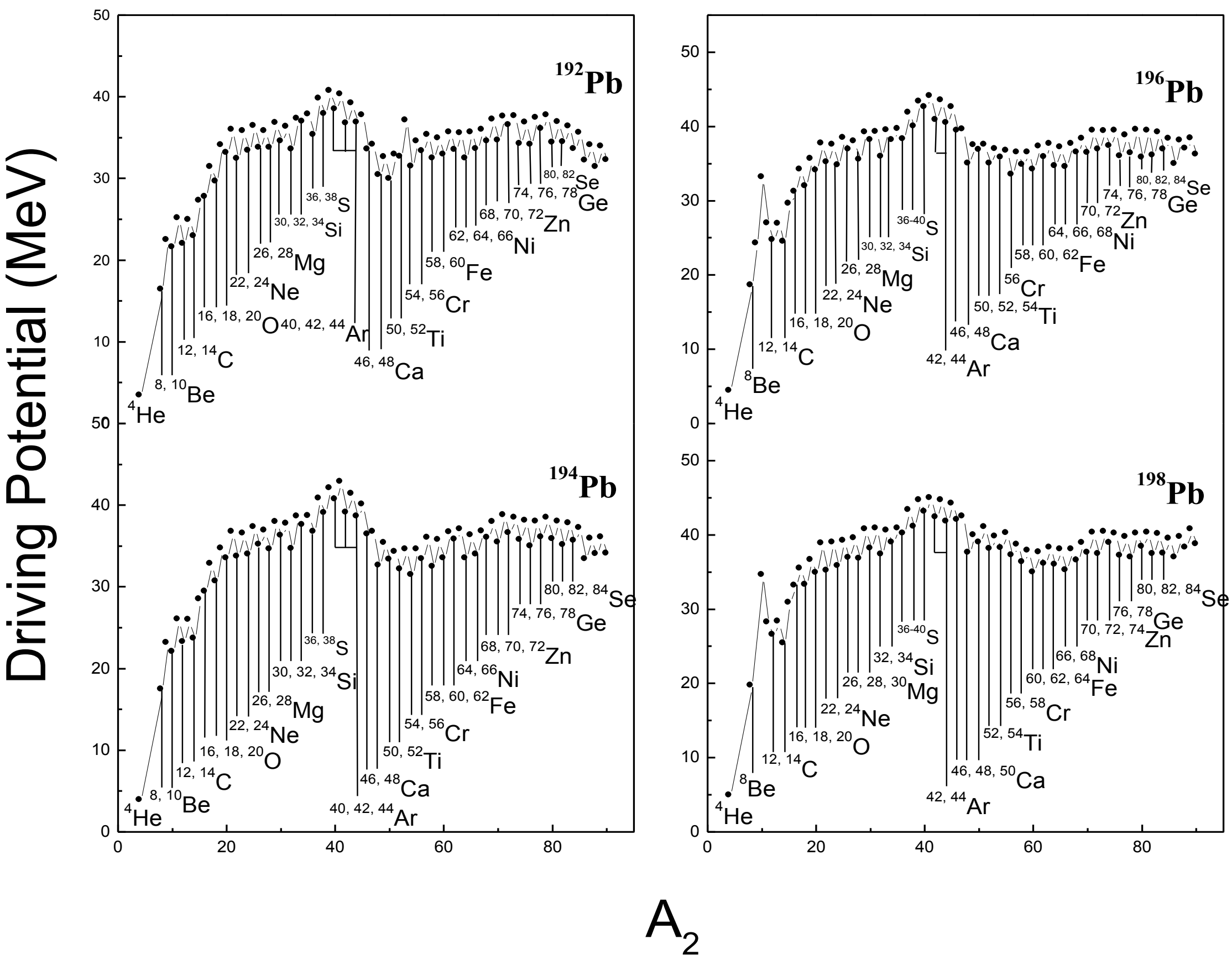

**Fig. 4.** The same as for Fig.1 but for Pb isotopes

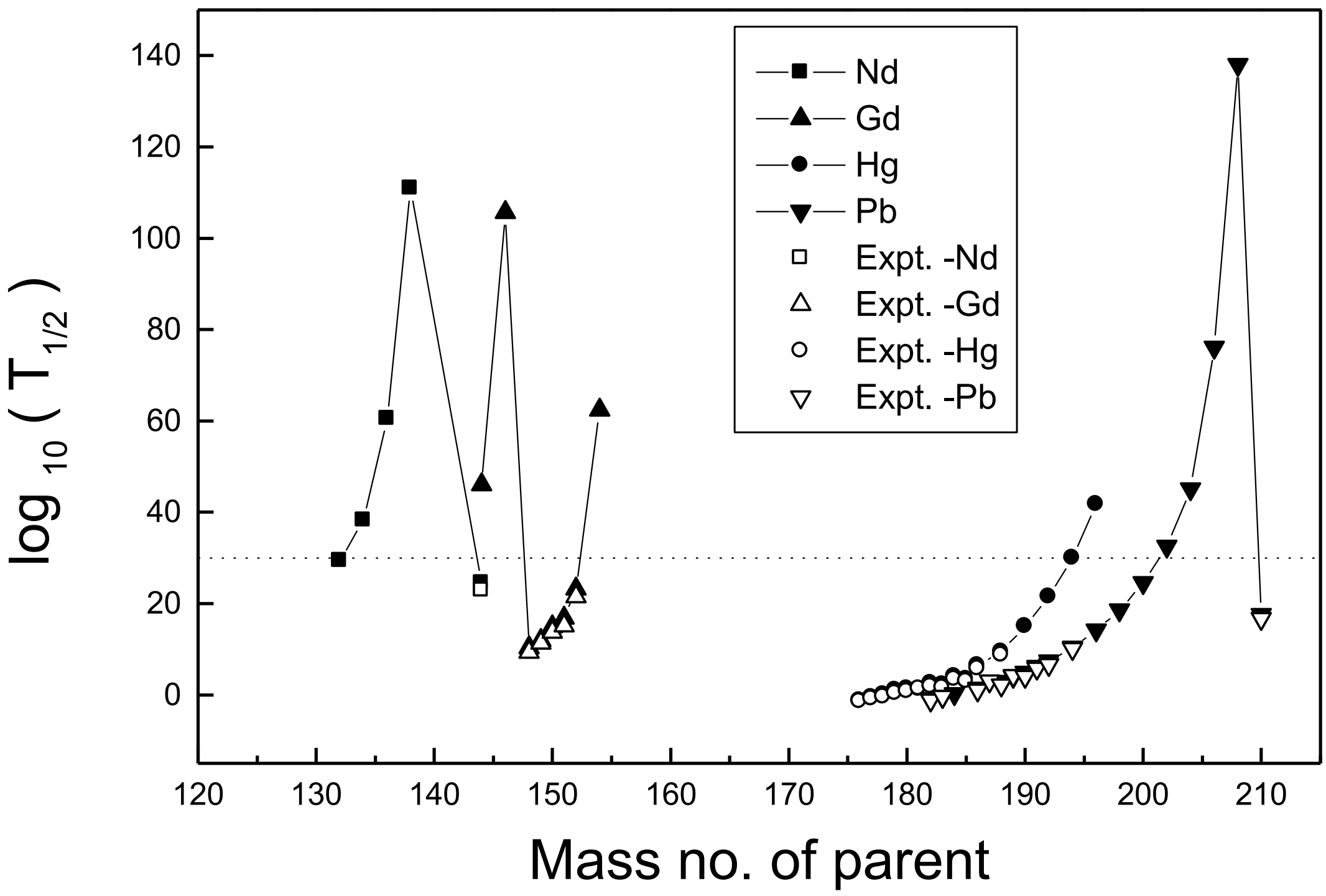

**Fig. 5.** The logarithms of half lives for alpha emission from the chosen Nd, Gd, Hg and Pb parents calculated using CPPM and compared with the available experimental values.

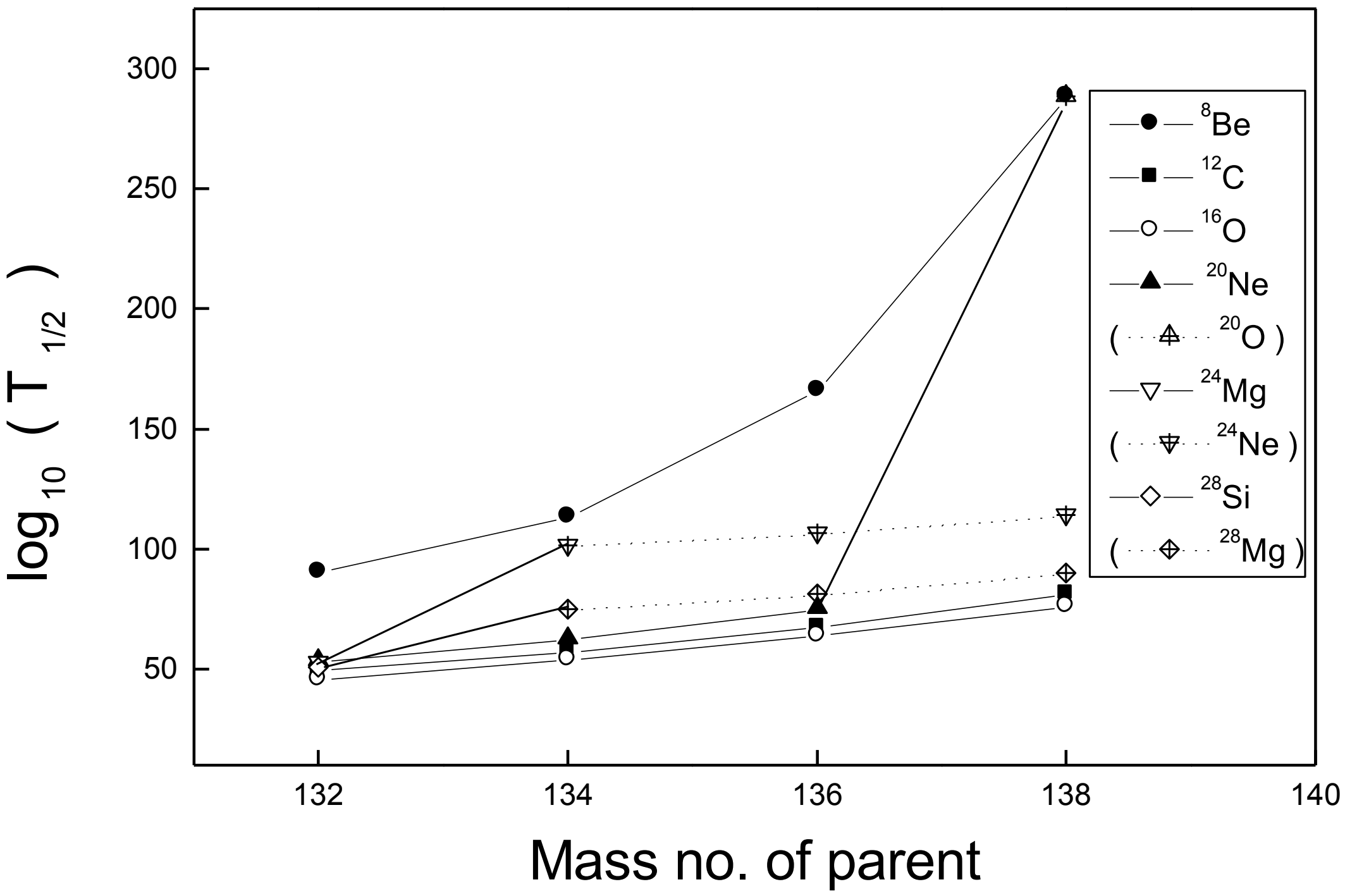

**Fig. 6.** $\log_{10}(T_{1/2})$ values for the N = Z clusters from Nd isotopes plotted as a function of the parent mass number.

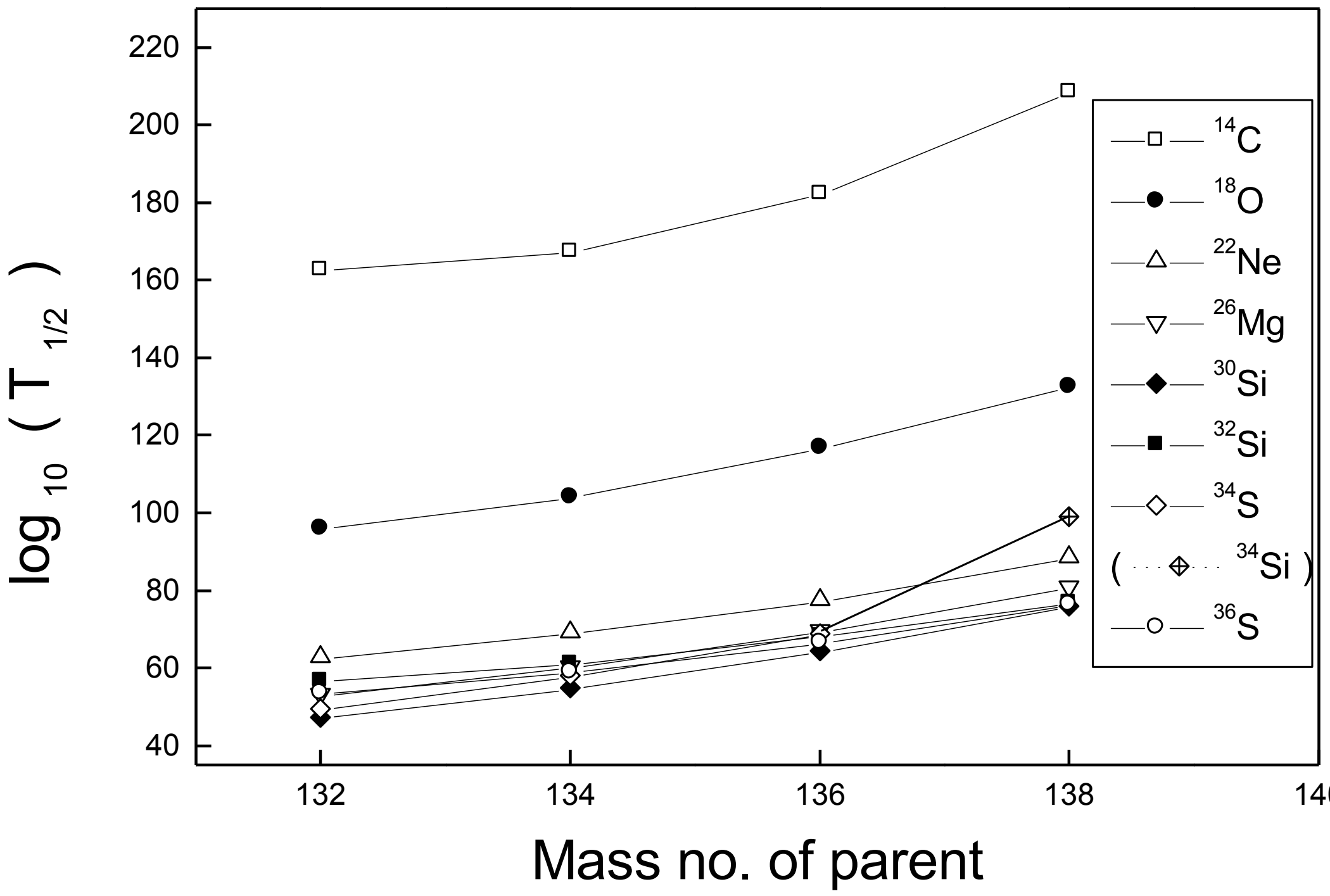

**Fig. 7.** The same as for Fig.6 but for N ≠ Z clusters from Nd isotopes.

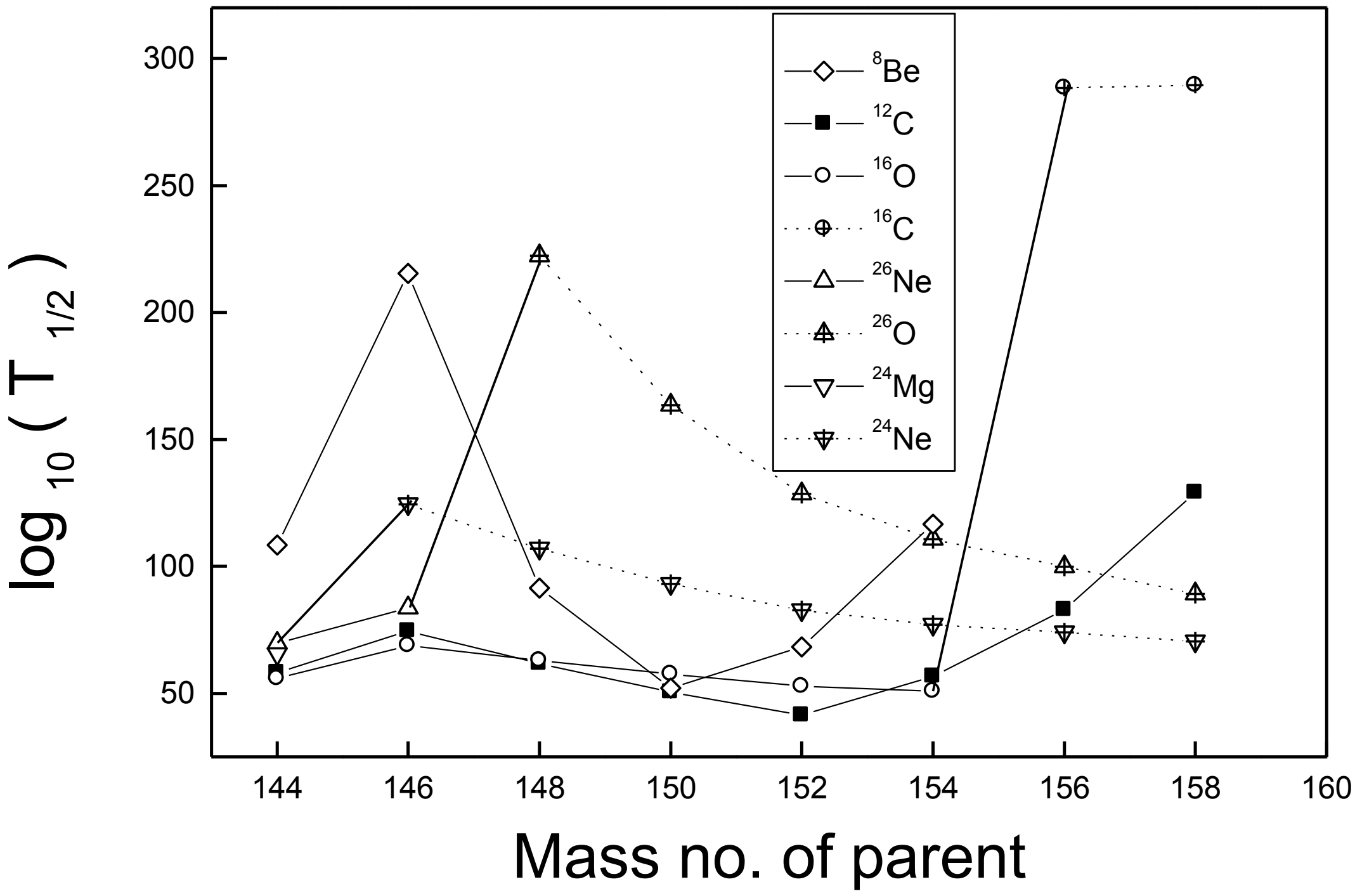

**Fig. 8.** $\log_{10}$ ($T_{1/2}$) values for the N = Z clusters from Gd isotopes plotted as a function of the parent mass number.

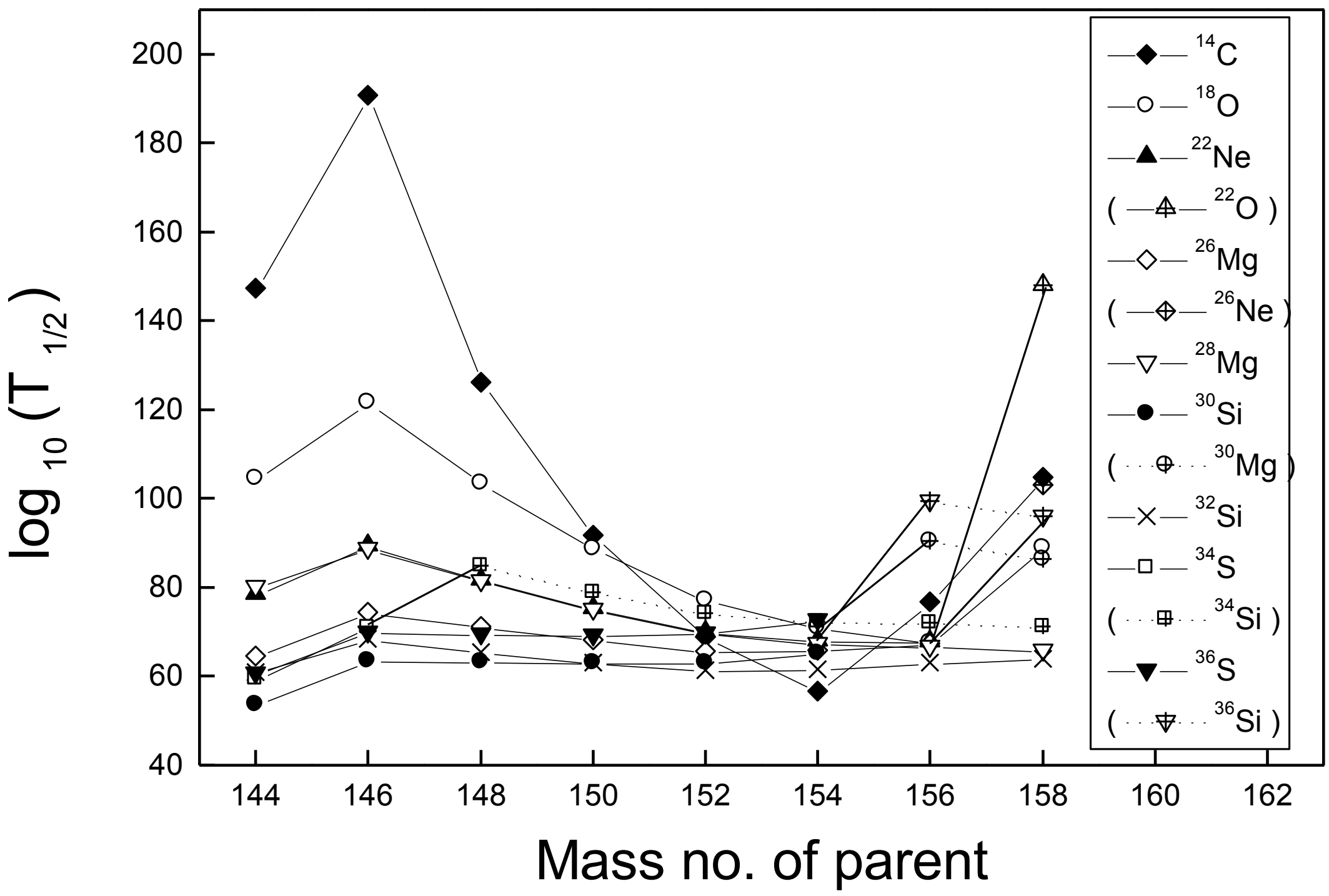

**Fig. 9.** The same as for Fig.8 but for N ≠ Z clusters from Gd isotopes.

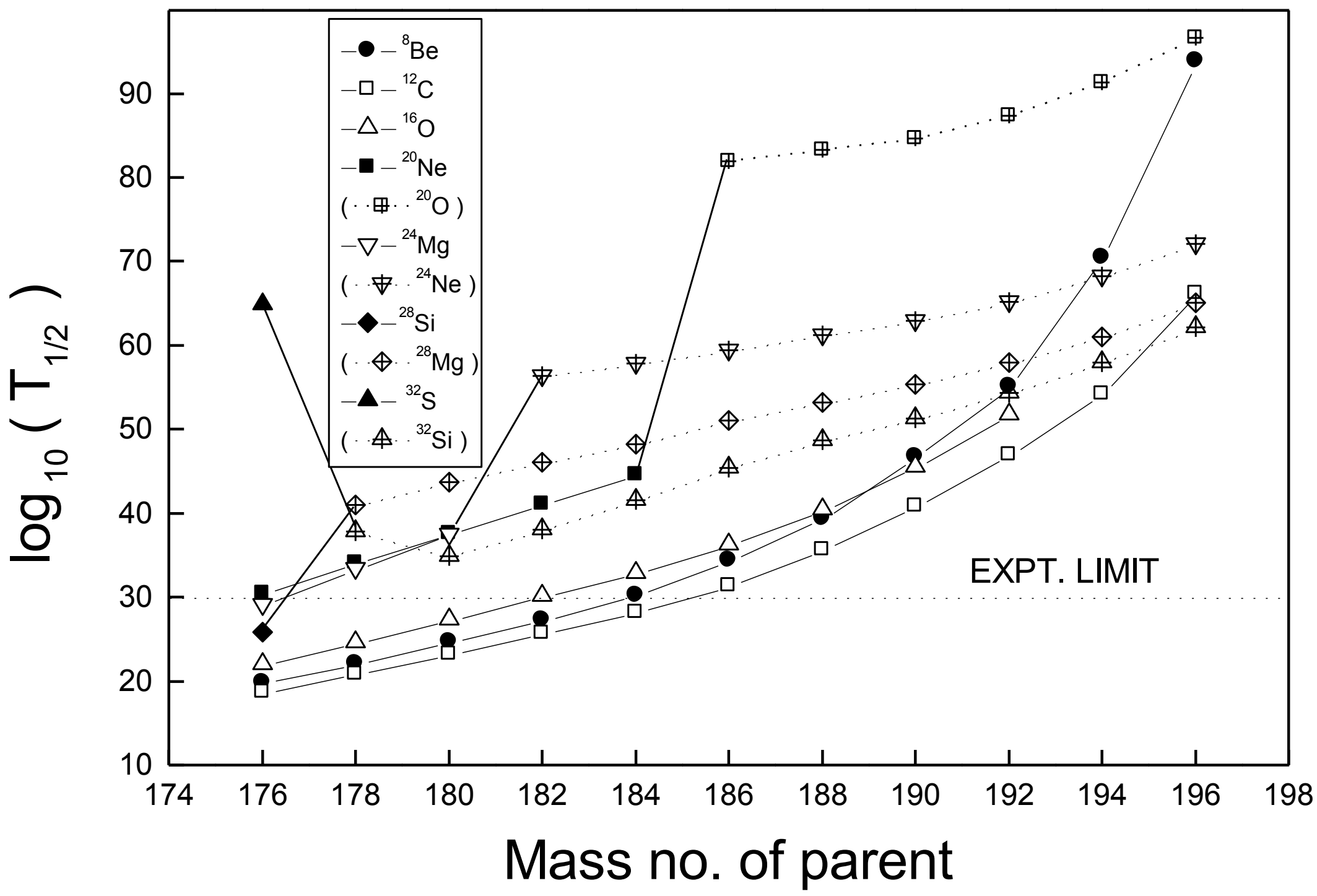

**Fig. 10.** $\log_{10}(T_{1/2})$ values for the N = Z clusters from Hg isotopes plotted as a function of the parent mass number.

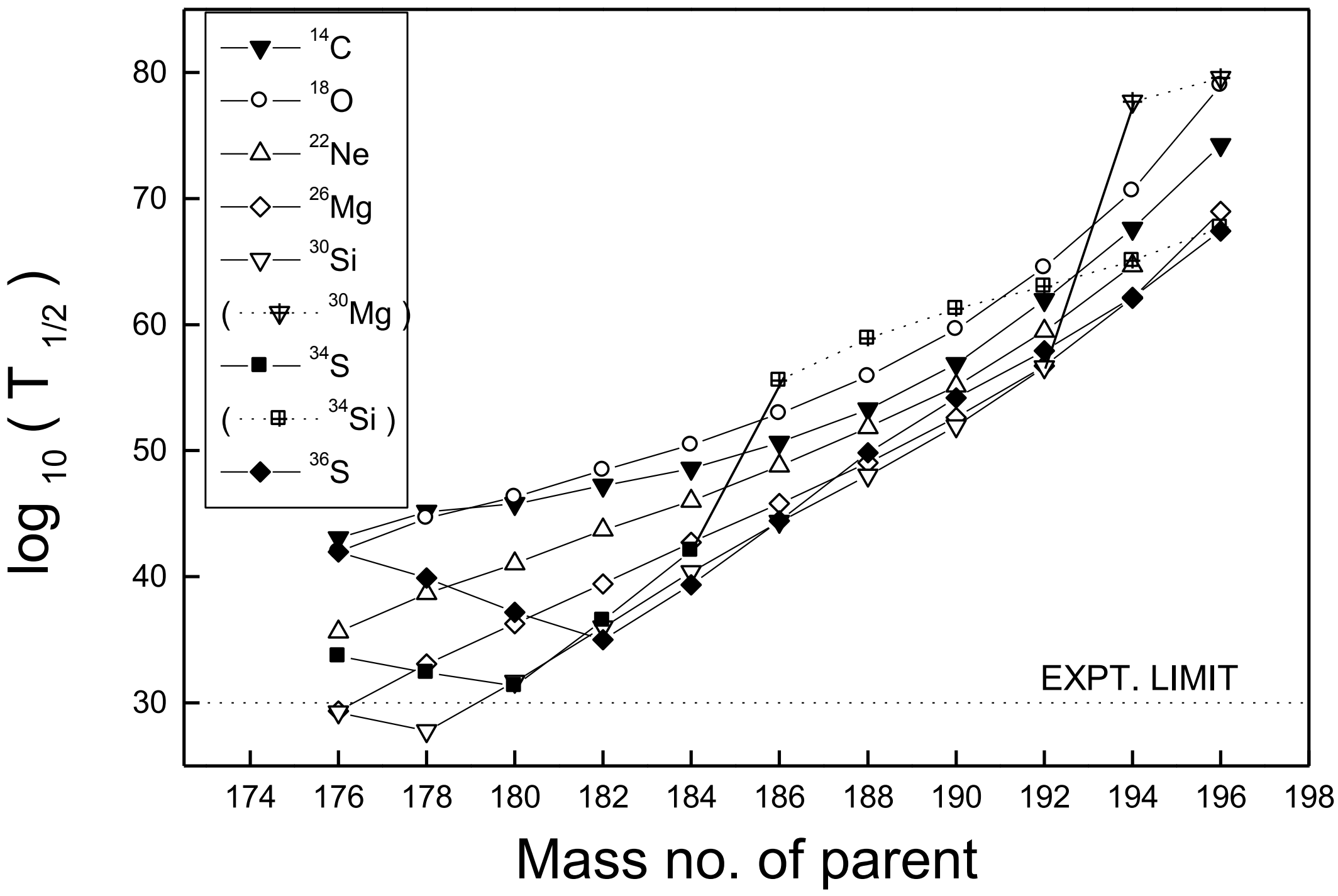

**Fig. 11.** The same as for Fig.10 but for N ≠ Z clusters from Hg isotopes.

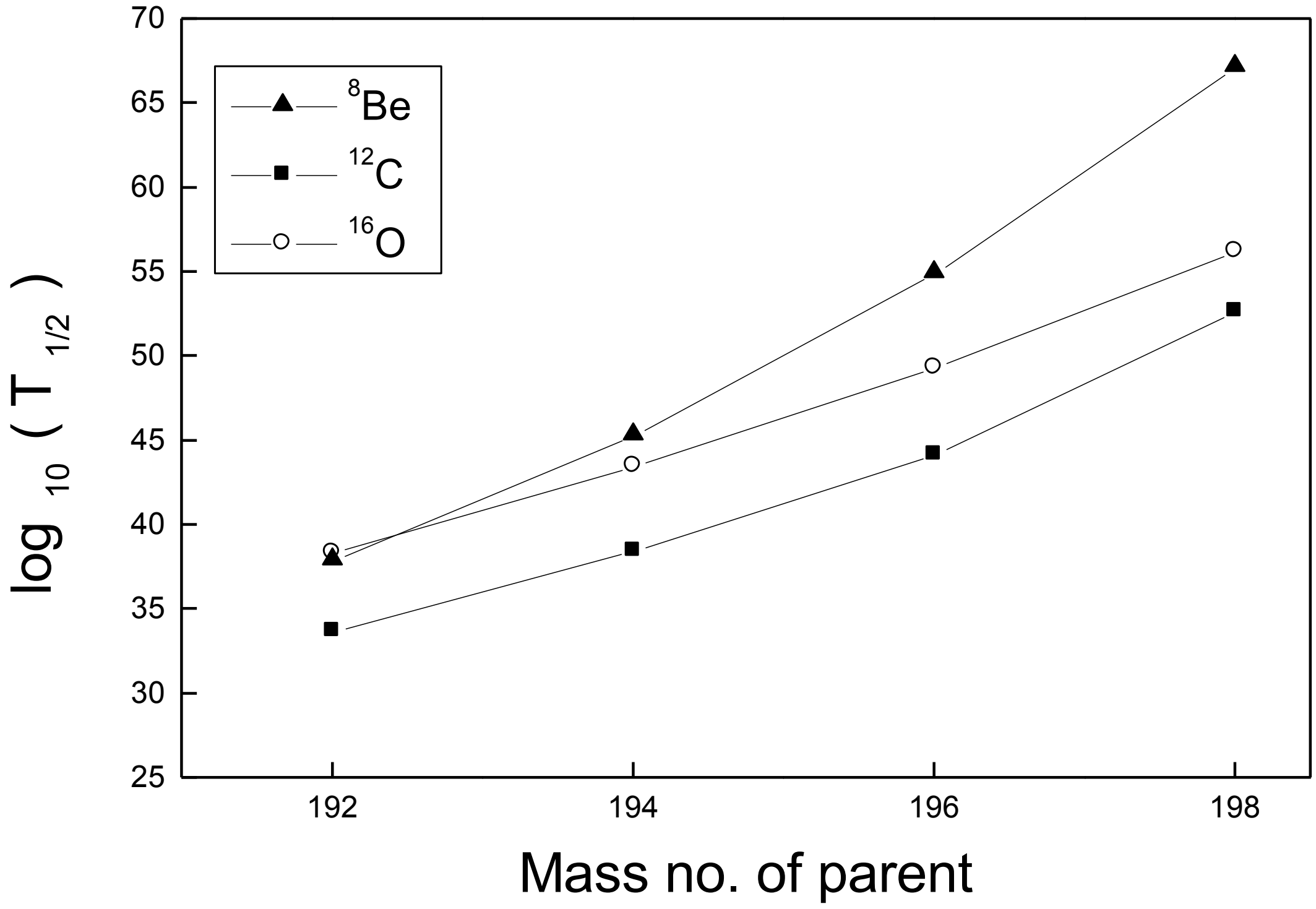

**Fig. 12.** $\log_{10}(T_{1/2})$ values for the N = Z clusters from Pb isotopes plotted as a function of the parent mass number.

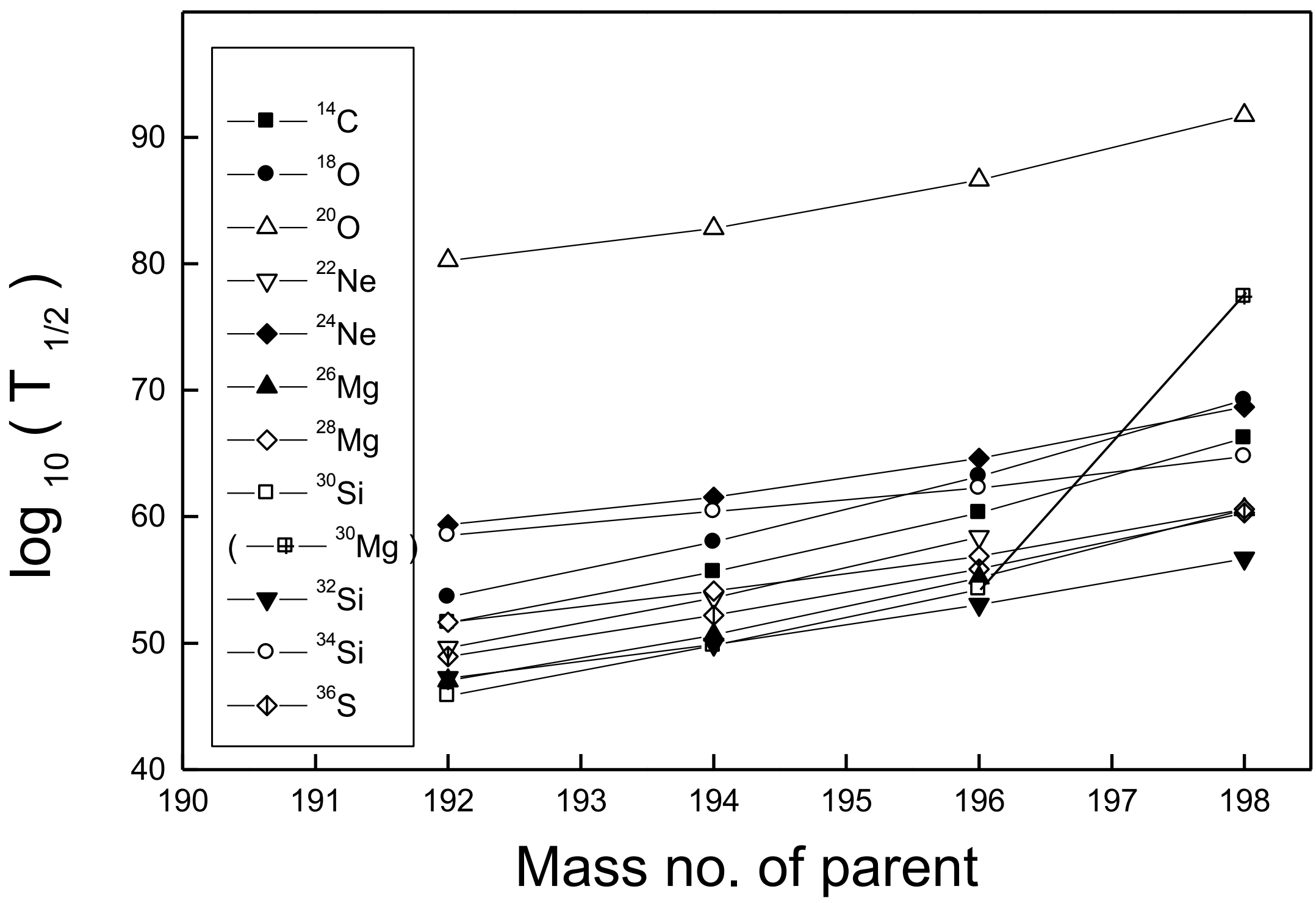

**Fig. 13.** The same as for Fig.10 but for N ≠ Z clusters from Pb isotopes.

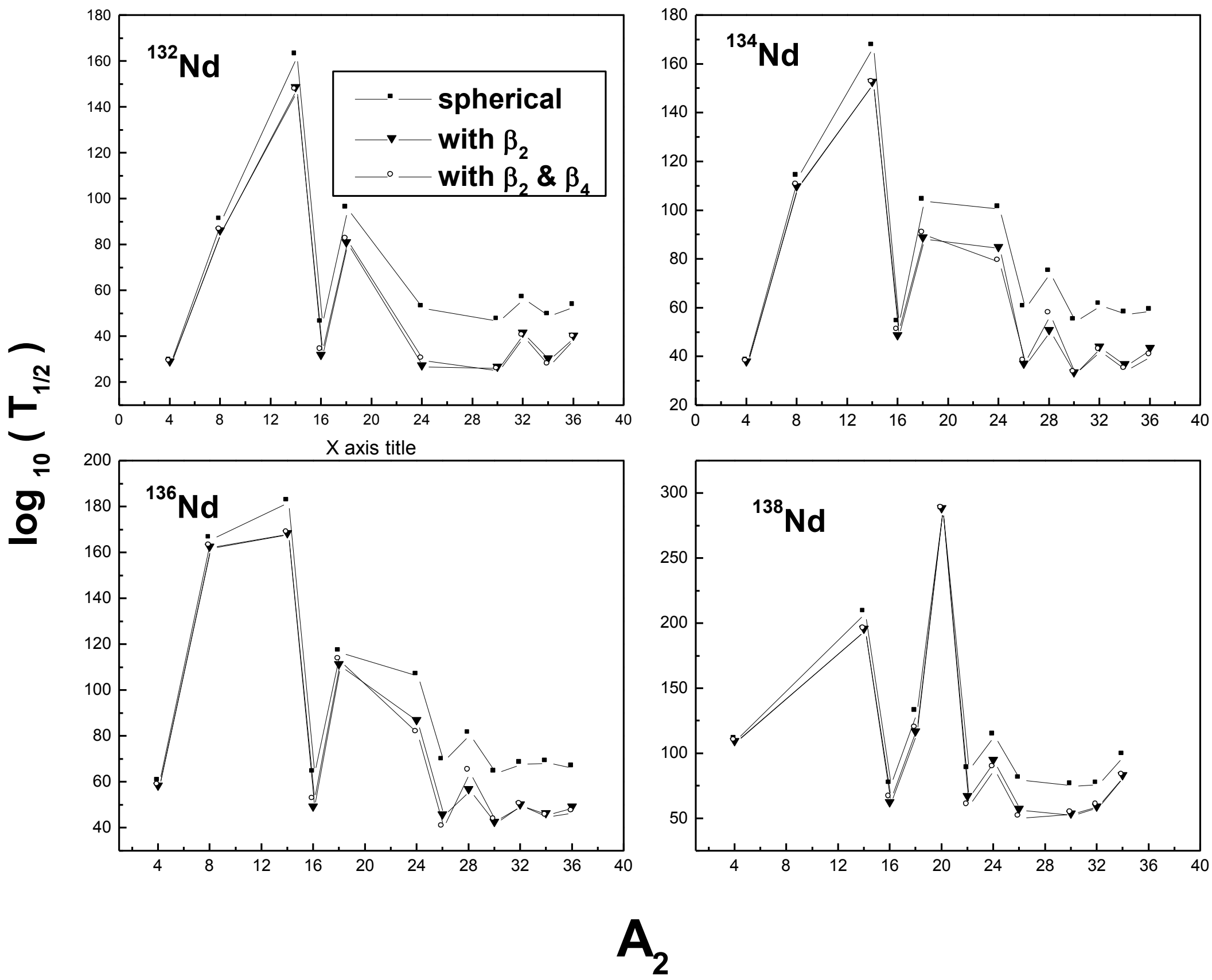

**Fig. 14.** Comparison of half life times for various clusters from $^{132-138}$Nd parents with and without including deformations.

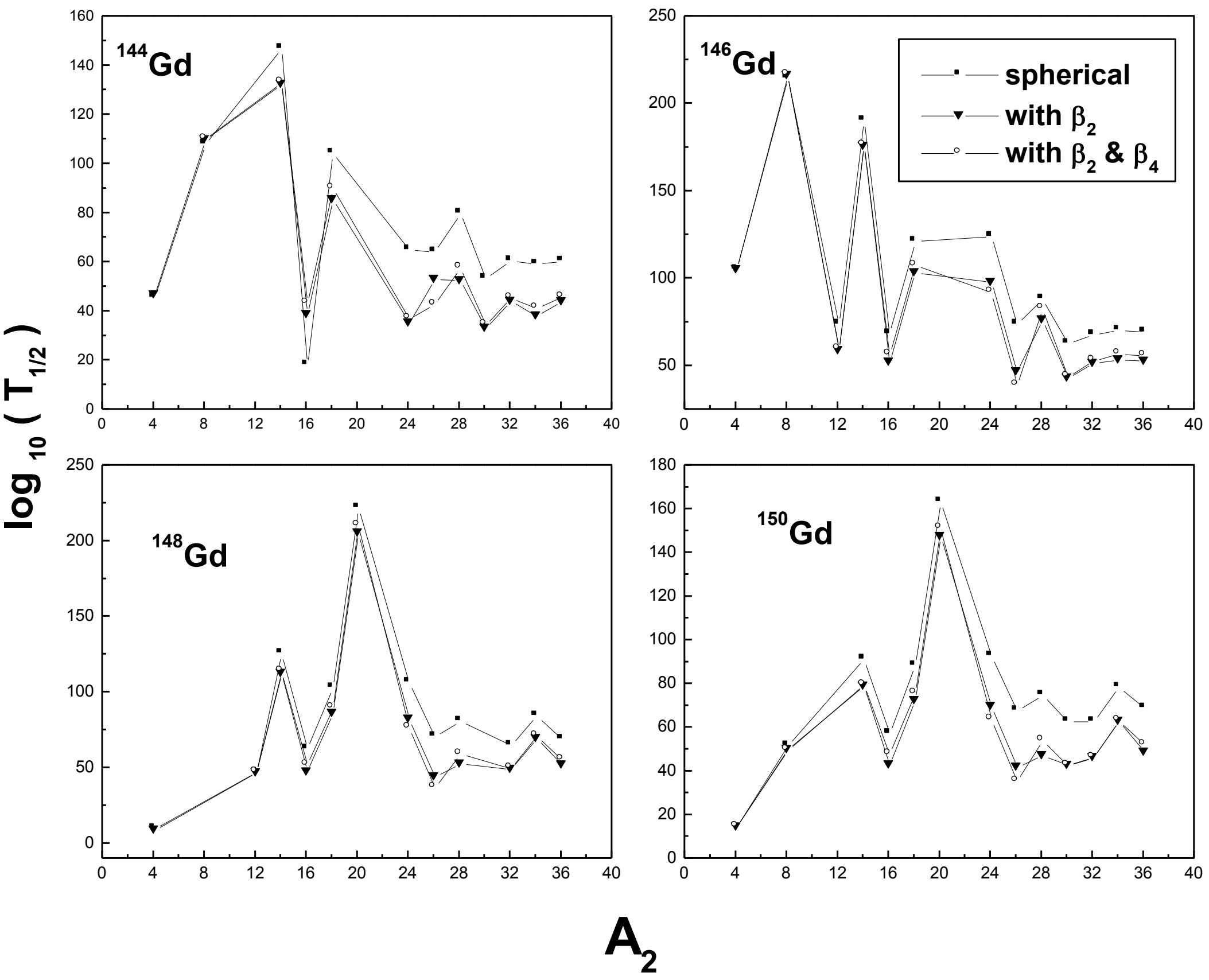

**Fig. 15.** The same as for Fig. 14 but for $^{144\text{-}150}$Gd parents

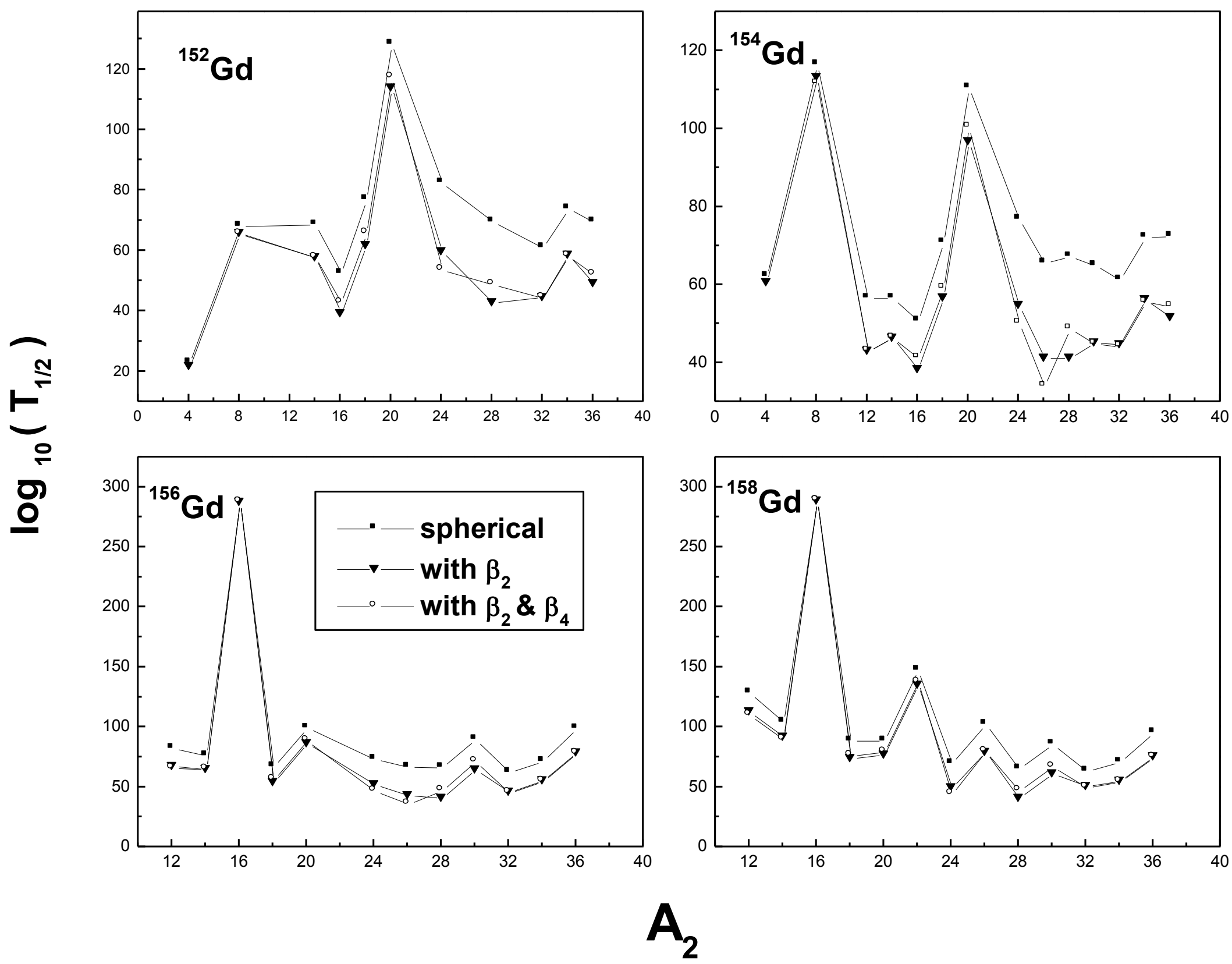

**Fig. 16.** The same as for Fig. 14 but for $^{152\text{-}158}$Gd parents

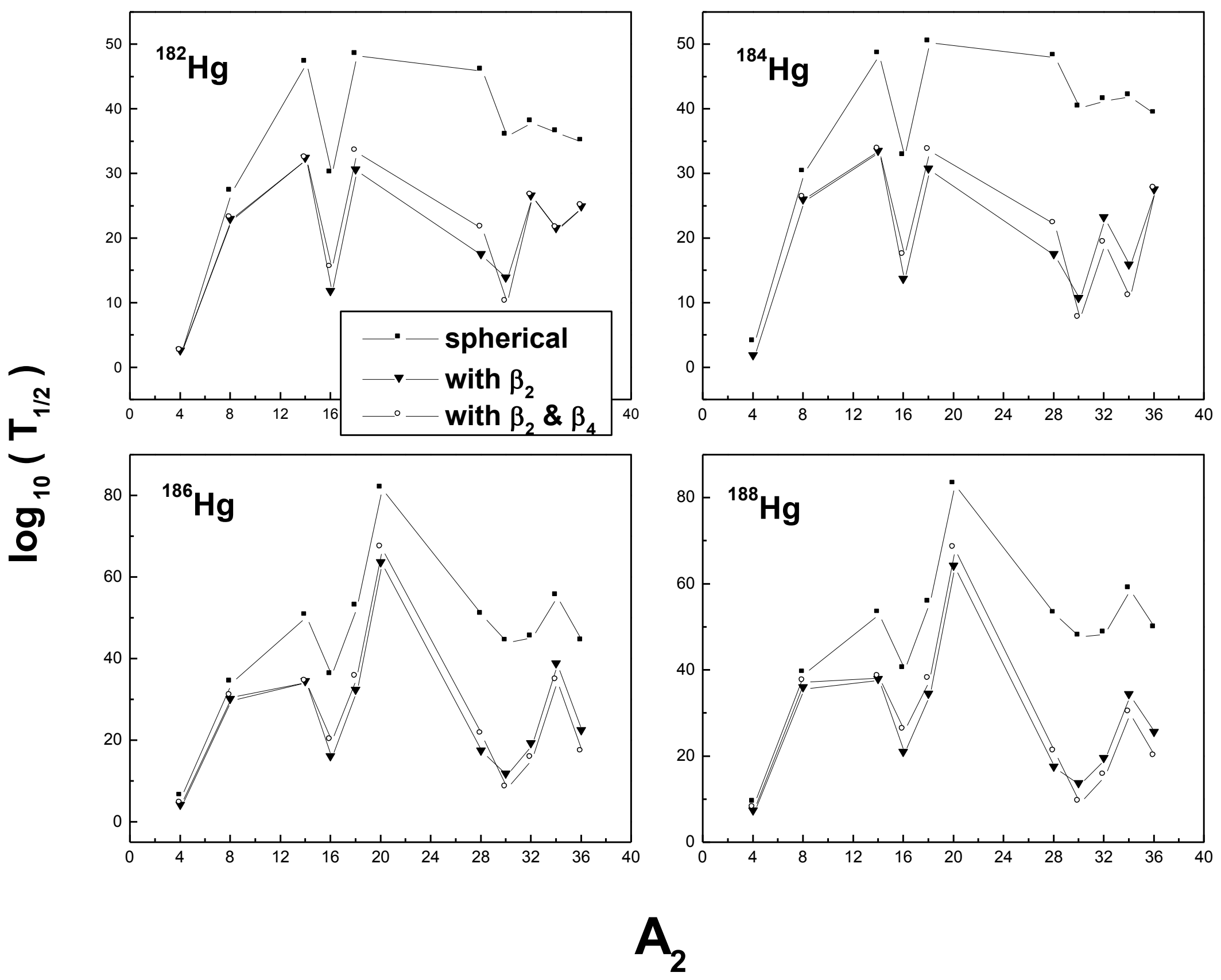

**Fig. 17.** The same as for Fig. 14 but for $^{182\text{-}188}$Hg parents

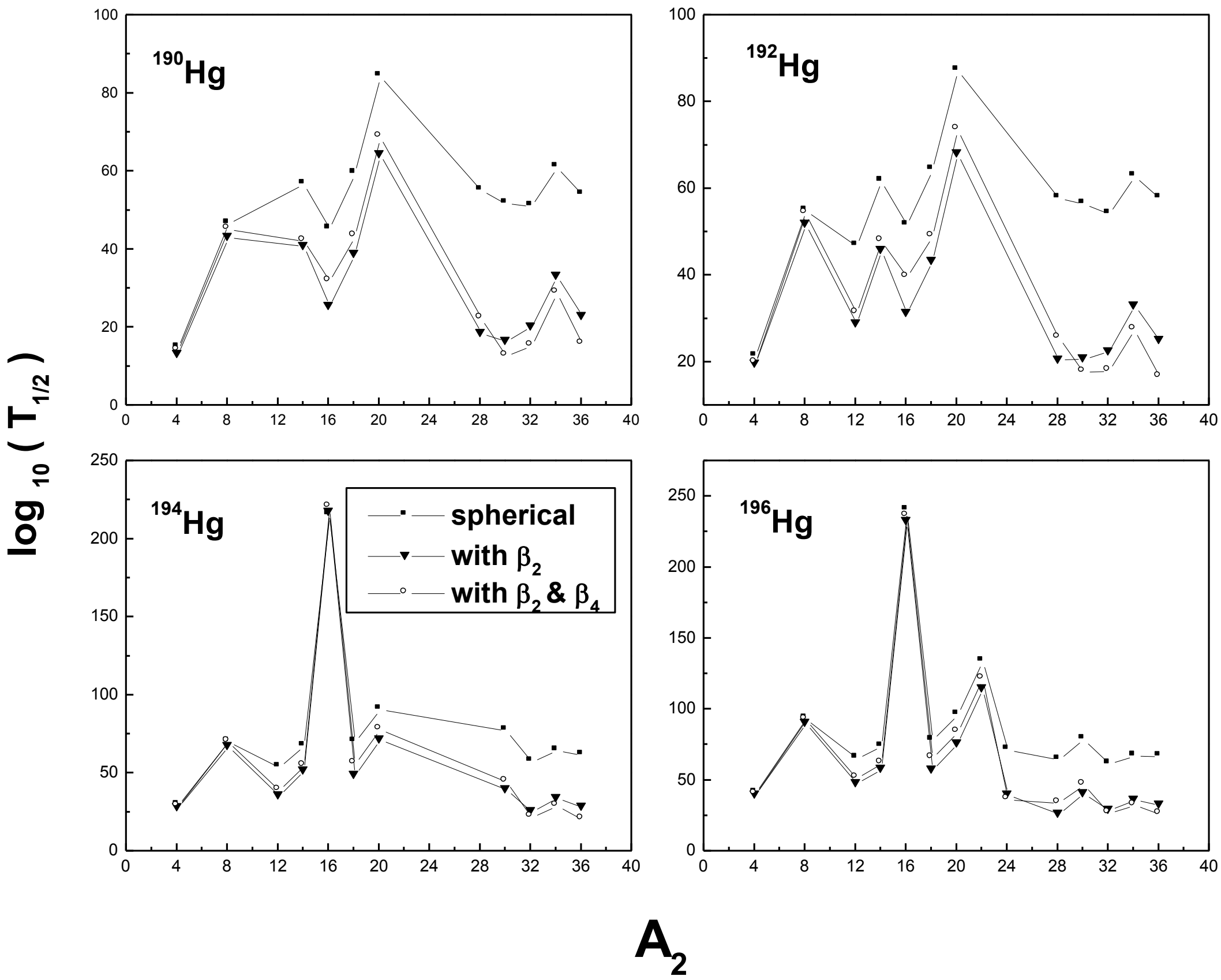

**Fig. 18.** The same as for Fig. 14 but for $^{190\text{-}196}$Hg parents

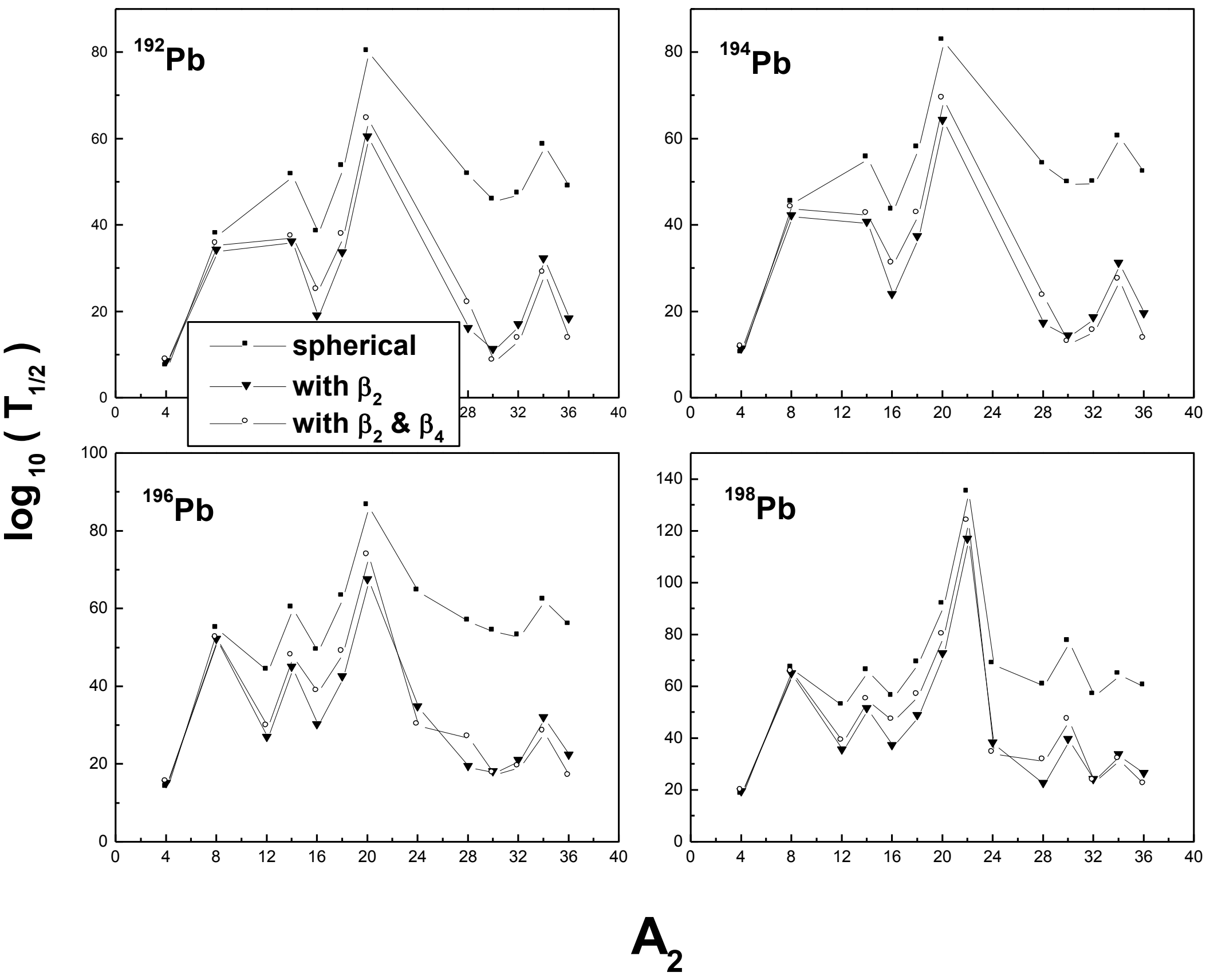

**Fig. 19.** The same as for Fig. 14 but for $^{192-198}$Pb parents.